\documentclass[aip,jcp,float,amsmath,amssymb,amsfonts,nofootinbib,superscriptaddress,reprint,twocolumn]{revtex4-1}
\usepackage{graphicx}
\usepackage{soul}
\usepackage[usenames,dvipsnames]{color}
\usepackage{hyperref}   % use for hypertext links, including those to external documents and URLs
\hypersetup{
	colorlinks = True,
	allcolors = {blue}
}

\begin{document}
\title{Thermofield Theory for Finite-Temperature Quantum Chemistry}
\author{Gaurav Harsha}
\affiliation{Department of Physics and Astronomy, Rice University, Houston, TX 77005-1892}

\author{Thomas M. Henderson}
\affiliation{Department of Physics and Astronomy, Rice University, Houston, TX 77005-1892}
\affiliation{Department of Chemistry, Rice University, Houston, TX 77005-1892}

\author{Gustavo E. Scuseria}
\affiliation{Department of Physics and Astronomy, Rice University, Houston, TX 77005-1892}
\affiliation{Department of Chemistry, Rice University, Houston, TX 77005-1892}

\date{\today}

\begin{abstract}
    Thermofield dynamics has proven to be a very useful theory in high-energy physics, particularly since it permits the treatment of both time- and temperature-dependence on an equal footing. We here show that it also has an excellent potential for studying thermal properties of electronic systems in physics and chemistry. We describe a general framework for constructing finite temperature correlated wave function methods typical of ground state methods. We then introduce two distinct approaches to the resulting imaginary time Schr\"{o}dinger equation, which we refer to as fixed-reference and covariant methods. As an example, we derive the two corresponding versions of thermal configuration interaction theory, and apply them to the Hubbard model, while comparing with exact benchmark results.
\end{abstract}

\maketitle

\section{Introduction
\label{sec1}}
The Schr\"{o}dinger equation for an interacting many-electron system can almost never be solved exactly due to the exponential scaling of the Hilbert space.\cite{dirac_quantum_1929} Over the years, several methods have been proposed to compute approximate solutions and properties for such systems. These methods (perhaps with the exception of density functional theory) fall into two broad categories: the path integral approach and the wave function approach.  The path integral formulation, generally used in condensed matter physics, is particularly useful when thermodynamically large systems can be approximated using an effective mean-field theory in the presence of perturbing interactions.  On the other hand, wave function methods (such as configuration interaction (CI), coupled cluster\cite{crawford_introduction_2000,bartlett_coupled-cluster_2007} (CC), or matrix product states\cite{ostlund_thermodynamic_1995,dukelsky_equivalence_1998} (MPS)) are often preferred for the study of finite many-body systems such as atoms and molecules -- systems generally of interest in chemistry.

While the path integral formulation can be naturally extended to study the thermal behaviour of a system via Matsubara's imaginary time formalism,\cite{matsubara_new_1955} the usage of typical wave function methods breaks down at non-zero temperatures.  This is because at non-zero temperatures, a quantum system is described by an ensemble density matrix, and one would need to solve for the entire spectrum of the Hamiltonian, as opposed to finding just the ground state energy and the corresponding wave function at zero temperature.  
There are a number of reasons for developing finite temperature quantum chemistry, a subject that has recently attracted considerable interest.\cite{verstraete_matrix_2004,feiguin_finite-temperature_2005, white_minimally_2009, stoudenmire_minimally_2010, pizorn_real_2014, hermes_finite_temperature_2015, czarnik_variational_2016, santra_finite-temperature_2017, zgid_finite_2017, claes_finite-temperature_2017, white_time-dependent_2018, hummel_finite_2018, hirata_converging_2018}
The finite temperature variants of the coupled cluster method\cite{hummel_finite_2018,white_time-dependent_2018} are of particular relevance to our present work.
These techniques are based on an extension of the time-dependent CC method to imaginary time, and are framed loosely along similar lines as the Thermal Cluster Cumulant theory proposed by Mukherjee \textit{et.al.},\cite{sanyal_thermal_1992,sanyal_systematic_1993,mandal_thermal_1998,mandal_finite_temperature_2003} which uses a finite-temperature generalization of Wick's theorem to introduce a thermal normal ordering. The interaction-picture Boltzmann operator is then compactly expressed as a thermal normal ordered exponential of some cluster operators and a number. Such an exponential ansatz is naturally size-extensive and forms a thermal analogue of ground-state CC.

Umezawa \textit{et. al.} proposed an alternative approach, known as thermofield dynamics,\cite{matsumoto_thermo_1983,semenoff_functional_1983,umezawa_methods_1984,evans_heisenberg_1992} which provides a real-time approach as opposed to Matsubara's imaginary time formalism.  Thermofield dynamics (TFD) can be conveniently described in an operator or wave function formulation, and was originally proposed as a method to study time-dependent and non-equilibrium phenomena in many-body quantum systems, something which cannot be accomplished easily within the original Matsubara formalism, but is addressed within the Keldysh formalism.\cite{keldysh_diagram_1965} TFD has been used widely in high energy physics\cite{maldacena_eternal_2003,shenker_multiple_2014,israel_thermo-field_1976,yang_complexity_2018} as well as for the study of time-dependent and open quantum systems.\cite{suzuki_thermo_1985,de_vega_thermofield-based_2015,borrelli_quantum_2016,chen_finite_2017,borrelli_simulation_2017}

The wavefunction prescription of TFD has tremendous potential for studying thermal properties of quantum many body systems in physics and chemistry. However, its application to compute equilibrium thermal properties has been relatively unexplored and generally limited to the mean-field level.\cite{hatsuda_mean_1989,walet_thermal_1990} Nonetheless, there are some methods such as Ancilla Density Matrix Renormalization Group\cite{feiguin_finite-temperature_2005,nocera_symmetry-conserving_2016} which try to harness the novel features of TFD. In this paper, we present a general framework based on TFD to construct wave function methods, extending traditional ground-state methods, to compute thermal averages of physical quantities. Our TFD-based approach provides an alternative to existing techniques.\cite{verstraete_matrix_2004,feiguin_finite-temperature_2005, white_minimally_2009, stoudenmire_minimally_2010, pizorn_real_2014, hermes_finite_temperature_2015, czarnik_variational_2016, santra_finite-temperature_2017, zgid_finite_2017, claes_finite-temperature_2017, white_time-dependent_2018, hummel_finite_2018, hirata_converging_2018,sanyal_thermal_1992,sanyal_systematic_1993,mandal_thermal_1998,mandal_finite_temperature_2003} While it shares similarities with many of these techniques (albeit from a different perspective), detailed analysis needs to be carried out to understand the precise distinctions between the different approaches.
 
\section{Thermofield Dynamics
\label{sec2}}
At non-zero temperatures, the expectation value of an operator $A$ (i.e. its thermal average) must be evaluated as an ensemble average
\begin{align}
\langle A \rangle_\beta
 &= \frac{1}{\mathcal{Z}} \mathrm{Tr}(A \hat{\rho}),
\label{thermal-trace}
\\
 &= \frac{1}{\mathcal{Z}} \sum_m \left \langle m \left \rvert~ A \mathrm{e}^{-\beta H } ~\right \lvert m \right \rangle
\label{ensemble-avg},
\end{align}
where $\beta$ is the inverse temperature, $H$ is the Hamiltonian, $\{ \lvert m\rangle \}$ forms a complete orthonormal basis and the partition function $\mathcal{Z}$ is defined as the trace of $\hat{\rho}$, the density operator
\begin{equation}
\hat{\rho} = \mathrm{e}^{-\beta H}, \quad \mathcal{Z} = \sum_m \left \langle m \left \rvert ~ \hat{\rho} ~\right \lvert m \right \rangle.
\end{equation}
One can choose to work within the grand canonical or the canonical ensemble by appropriately defining the Hamiltonian $H$ with or without a chemical potential term.  In this paper, unless mentioned otherwise, we shall work in the grand canonical ensemble.

The central idea in TFD is to express the ensemble average of an operator $A$ in Eq.~\ref{ensemble-avg} as an expectation value over just one state $\lvert \Psi(\beta) \rangle$, known as the \textit{thermofield double state}, \textit{thermal vacuum} or just \textit{thermal state}:
\begin{equation}
\langle \hat{A} \rangle_\beta = \frac{\langle \Psi(\beta) \rvert~ \hat{A} ~ \lvert \Psi (\beta) \rangle}{\langle \Psi(\beta) ~\rvert ~\Psi(\beta) \rangle}.
\label{thermal-exp}
\end{equation}
This idea is realized by introducing a copy of the original Hilbert space $\mathcal{H}$, known as the auxiliary or \textit{tilde}-conjugate space $\tilde{\mathcal{H}}$, such that:
\begin{enumerate}
\item For every state $\lvert \psi \rangle$ in $\mathcal{H}$, there is a copy $\lvert \tilde{\psi}\rangle$ in $\tilde{\mathcal{H}}$ and likewise for operators.

\item The \textit{tilde} operators obey similar (anti-) commutation rules as their un-tilde counterparts. For instance, for a bosonic (fermionic) spin-orbital $k$, we have the extended set of field operators
\[
c_k,\: c^\dagger_k;\: \tilde{c}_k,\: \tilde{c}^\dagger_k
\]
following the (anti-) commutation rules given by
\begin{subequations}
    \label{comm12}
    \begin{align}
        [ c_k, c^\dagger_k ]_{\mp} &= 1 = [\tilde{c}_k, \tilde{c}^\dagger_k]_{\mp},
        \\
        [ c_k, \tilde{c}_k ]_{\mp} &= 0 = [c^\dagger_k, \tilde{c}^\dagger_k]_{\mp},
    \end{align}
\end{subequations}
where the convention for commutator / anti-commutator is defined as
\begin{equation}
[ A, B] _{-\eta} = A B - \eta B A,
\end{equation}
such that $\eta = -1$ in Eq.~\ref{comm12} produces the anticommutation rules for fermions and $\eta = +1$ produces the commutation rules for bosons.
        
\item A tilde conjugation operation transforms operators between $\mathcal{H}$ and $\tilde{\mathcal{H}}$ with the following general rules:
\begin{align}
\widetilde{\left ( \tilde{c}_k \right) } &= \eta c_k,
\label{conj2}
\\
\widetilde{\left ( \alpha c_k + \delta c^\dagger_q \right )} &= \alpha^\star \tilde{c}_k + \delta^\star \tilde{c}^\dagger_q,
\\
\widetilde{\left ( c_k c^\dagger_q \right )} &= \tilde{c}_k \tilde{c}^\dagger_q, 
\end{align}
where $\alpha^\star, \delta^\star$ are complex conjugates of $\alpha, \delta$ respectively, and $\eta = \pm 1$ for bosons / fermions in Eq.~\ref{conj2}.
With these conjugation rules, a Hamiltonian for the \textit{tilde} system can be defined, generally denoted by $\tilde{H}$.

\item The time-dependent Schr\"{o}dinger equation in $\tilde{\mathcal{H}}$ becomes ($\hbar = 1$)
\begin{equation}
-\mathrm{i} \frac{\partial }{\partial t} \left \lvert \psi \right \rangle =  \tilde{H} \left \lvert \psi \right \rangle.
\end{equation}

\item Operators in the physical space do not act on states in the tilde space, and \textit{vice versa}.
\end{enumerate}
The conjugation and doubling of the Hilbert space has well-justified connections with Hopf algebra.\cite{celeghini_thermo_1998,floquet_lie_2017}

In the expanded space, the thermal state in Eq.~\ref{thermal-exp} can be expressed as
\begin{equation}
\left | \Psi (\beta) \right\rangle = \mathrm{e}^{-\beta H/2} \sum_m |m\rangle \otimes |\tilde{m}\rangle,
\label{fci-vac}
\end{equation}
where $\{|m\rangle\}$ forms an orthonormal basis in $\mathcal{H}$ and $H$ is the Hamiltonian of the system (and so does not act on states in the tilde space). The norm of the thermal state gives the partition function
\begin{equation}
\mathcal{Z} = \left \langle \Psi(\beta) \left \vert \right .\Psi(\beta) \right \rangle.
\label{part-fun}
\end{equation}
Note that at infinite temperature or $\beta=0$, the thermal state is merely given by
\begin{equation}
\left \vert \mathbb{I} \right \rangle = \left \vert \Psi (\beta=0) \right \rangle = \sum_m \vert m\rangle \otimes \vert \tilde{m}\rangle.
\end{equation}
Since the Hamiltonian is no longer relevant, the state $\vert \mathbb{I} \rangle$ only depends on the structure of the Hilbert space and can be computed exactly.  Consequently, $\vert \mathbb{I} \rangle$ is analogous to the identity operator and is independent of the choice of basis $\{\vert m \rangle \}$.

The  $|\tilde{m}\rangle$ states in Eq.~\ref{fci-vac} perform the role of a tracer.  That is, for some operator $\hat{A}$ which acts only on the physical states,
\begin{subequations}
\begin{align}
\langle \hat{A} \rangle_\beta 
 &= \frac{\langle \Psi(\beta) \rvert~ \hat{A} ~ \lvert \Psi (\beta) \rangle}{\langle \Psi(\beta) ~\rvert ~\Psi(\beta) \rangle}
\\
 &= \frac{1}{\mathcal{Z}} \, \sum_{m,n} \langle m,\tilde{m}| \mathrm{e}^{-\beta \, H/2} \, \hat{A} \, \mathrm{e}^{-\beta \, H/2} |n,\tilde{n}\rangle
\\
 &= \frac{1}{\mathcal{Z}} \, \sum_{m} \langle m| \mathrm{e}^{-\beta \, H/2} \, \hat{A} \, \mathrm{e}^{-\beta \, H/2} |m\rangle
\\
 &= \frac{1}{\mathcal{Z}} \, \mathrm{Tr}\left(\mathrm{e}^{-\beta \, H/2} \, \hat{A} \, \mathrm{e}^{-\beta \, H/2}\right)
\\
 &= \frac{1}{\mathcal{Z}} \, \mathrm{Tr}\left(\mathrm{e}^{-\beta \, H} \, \hat{A}\right)
\end{align}
\end{subequations}
where we have used the shorthand notation $|m,\tilde{m}\rangle = |m\rangle \otimes |\tilde{m}\rangle$.

Conventional applications of TFD have been mostly centered around the study of dynamics of quantum systems at finite temperatures which is governed by a new Hamiltonian
\begin{equation}
    H_\textrm{th}= H - \tilde{H},
\end{equation}
which drives the real-time dynamics of the thermal state through the Schr\"{o}dinger equation
\begin{equation}
    \mathrm{i} \frac{\partial}{\partial t} \vert \Psi (\beta) \rangle = H_\textrm{th} \vert \Psi (\beta) \rangle.
\end{equation}
This real-time Schr\"{o}dinger equation does not provide any prescription to compute the thermal state $\vert \Psi (\beta) \rangle$ and hence cannot be used to compute the partition function and other equilibrium thermal properties for the original system.
Instead, we realize that $\vert \Psi(\beta)\rangle$, while not an eigenstate of the physical Hamiltonian $H$, obeys an imaginary time Schr\"{o}dinger equation, given by
\begin{equation}
\frac{\partial}{\partial \beta} \left \lvert \Psi(\beta)\right \rangle = -\frac{1}{2} H \left \lvert \Psi(\beta) \right \rangle,
\label{imag-schro}
\end{equation}
which allows us to construct the thermal state at any $\beta$ by integrating Eq.~\ref{imag-schro} starting from temperature $\beta_0$ where the thermal state $\vert \Psi (\beta_0) \rangle$ is known exactly.
Similarly, if an explicit dependence on the chemical potential $\mu$ (or rather $\alpha = \beta \mu$) is considered, an equation for the evolution in $\alpha$ can be established, i.e.
\begin{align}
\vert \Psi(\alpha, \beta) \rangle &= \mathrm{e}^{(\alpha N -\beta H)/2} \vert \mathbb{I} \rangle,
\label{beta-chem-th-state}
\\
\frac{\partial}{\partial \alpha} \vert \Psi (\alpha, \beta) \rangle &= \frac{1}{2} N \vert \Psi (\alpha, \beta) \rangle.
\label{mu-schro}
\end{align}
In what follows, we explicitly describe the $\beta$ evolution and provide a prescription to extend the formalism to the $\alpha$ evolution in accordance with Eq.~\ref{mu-schro}. For the sake of simplicity, we will refer to $\alpha$ as the chemical potential.

\subsection{Thermal Mean-Field Theory
\label{sec2a}}
While Eq.~\ref{fci-vac} is formally exact, it is in practice impossible to determine the thermal state exactly except for the simplest model systems (it is, after all, a highly non-trivial problem to compute a single state and Eq.~\ref{fci-vac} requires us to compute every state).  Just as in ground state calculations, practical applications require a systematic way of approximating $\left \lvert \Psi (\beta)\right \rangle$.

The most elementary approximation one can invoke is the mean-field approach, wherein the Hamiltonian $H$ is approximated by a one-body mean-field Hamiltonian (e.g. the Fock operator), and correspondingly, the mean-field thermal vacuum takes the form
\begin{equation}
\left \lvert \Psi (\beta) \right \rangle_{\textrm{mf}} = \left \lvert 0 (\beta) \right \rangle = \mathrm{e}^{-\beta H_0/2} \sum_m |m\rangle \otimes |\tilde{m}\rangle,
\label{mf-vac}
\end{equation}
where $H_0$ is the mean-field Hamiltonian. This mean-field state satisfies the imaginary time Schr\"{o}dinger equation not for $H$ but for $H_0$:
\begin{equation}
\frac{\partial}{\partial \beta} |0(\beta)\rangle = -\frac{1}{2} \, H_0 \, |0(\beta)\rangle.
\label{mf-imag-schro}
\end{equation}
If we choose the states $\{ \lvert m \rangle \}$ to be the eigenstates of the mean-field Hamiltonian $H_0$, which can generally be computed without much computational effort, the thermal vacuum takes the form
\begin{equation}
\left \lvert 0(\beta) \right \rangle = \sum_m \mathrm{e}^{-\beta E_m/2}\vert m\rangle \otimes \vert \tilde{m}\rangle.
\end{equation}
The norm of $\left \lvert 0(\beta) \right \rangle$ gives the mean-field partition function
\begin{equation}
\mathcal{Z}_0 = \left \langle 0(\beta) \left \vert \right . 0(\beta) \right \rangle.
\label{mf-part-fun}
\end{equation}
Any quantity of interest, such as the energy, can then be approximately evaluated as an expectation value over $\left \lvert 0(\beta) \right \rangle$.

For fermions, working within the grand canonical formulation, the thermal state can also be expressed in terms of the single-particle Fock states:
\begin{equation}
\left \lvert 0(\beta) \right \rangle = \mathrm{e}^{-\beta H_0/2} \prod_{p \in \textrm{levels}} \Big (\vert 0\rangle_p \otimes \vert \tilde{0} \rangle_p + \vert 1\rangle_p \otimes \vert \tilde{1} \rangle_p\Big)
\end{equation}
where $|0\rangle_p$ and $|1\rangle_p$ respectively denote that the single-particle level $p$ is empty or occupied. If the levels $p$ are chosen to be the eigenstates of $H_0$, a simple thermal Bogoliubov transformation can be defined to construct the thermal field operators $\{ a_p(\beta), a_p^\dagger(\beta), \tilde{a}_p(\beta), \tilde{a}_p^\dagger(\beta) \}$ that create or annihilate quasi-particle excitations on to the thermal vacuum, i.e.
\begin{equation}
\begin{bmatrix} a_p(\beta) \\ \tilde{a}^\dagger_p(\beta) \end{bmatrix}
=
\begin{bmatrix} w_p & -z_p \\ z_p & w_p \end{bmatrix} \, \begin{bmatrix} c_p \\ \tilde{c}_p^\dagger \end{bmatrix},
\label{hfb-eq}
\end{equation}
such that
\begin{subequations}
\begin{align}
a_p |0(\beta)\rangle &= 0 = \tilde{a}_p |0 (\beta) \rangle,
\\
\langle 0(\beta) | a_p^\dagger  &= 0 = \langle 0(\beta) | \tilde{a}_p^\dagger.
\end{align}
\end{subequations}
It is easy to show that $w_p = \sqrt{1 - f_F(\epsilon_p)}$ and $z_p = \sqrt{f_F(\epsilon_p)}$, where $f_F$ is the Fermi-Dirac distribution function.  Here and in the following, we have dropped the explicit $\beta$-dependence in the operators $a_p$, $a_p^\dagger$, and their tilde counterparts.

While a similar transformation can be defined for bosons, henceforth we shall confine ourselves to the many-electron problem in this paper.  Furthermore, while such a transformation can also be built within the canonical ensemble, it is generally non-linear and introduces a new algebra for the thermal operators (see, e.g., the transformation for $SU(2)$ spin operators in Ref.~\onlinecite{suzuki_thermo_1986}).

The Bogoliubov transformation in Eq.~\ref{hfb-eq} allows us to form a physical intuition about the thermal operators: in the low $T$ (or high $\beta$) limit, the annihilation of a particle in the original space $\mathcal{H}$ is equivalent to creating a particle in the \textit{tilde}-space $\tilde{\mathcal{H}}$ and vice-versa, whereas in the high $T$ limit, these operators are completely mixed.

A general 2-body Hamiltonian, written in terms of the spin-orbital creation and annihilation operators as 
\begin{equation}
    H = \sum_{p,q} h_{pq} c_p^\dagger c_q + \frac{1}{4}\sum_{p,q} v_{pqrs} c_p^\dagger c_q^\dagger c_s c_r,
\end{equation}
can be expressed in terms of the thermal operators and takes the form
\begin{widetext}
\begin{align}
    H &= \sum_p z_p^2 \, \left(h_{pp} + \frac{1}{2} \, \sum_q z_q^2 \, v_{pqpq}\right) \nonumber \\
    &+ \sum_{pq} \left(h_{pq} + \sum_r z_r^2 \, v_{prqr}\right) \, \left(w_p \, w_q \, a_p^\dagger \, a_q - z_p \, z_q \, \tilde{a}_q^\dagger \, \tilde{a}_p + w_p \, z_q \, a^\dagger_p \, \tilde{a}^\dagger_q + w_q \, z_p \, \tilde{a}_p \, a_q\right) + \textrm{2-body terms}
\end{align}
\end{widetext}
Notice that the foregoing expression for the Hamiltonian is normal-ordered with respect to the thermal vacuum. The ordering we follow here is $a^\dagger \rightarrow \tilde{a}^\dagger \rightarrow \tilde{a} \rightarrow a$.
The thermal average of $H$ at the mean-field level is simply found by extracting the scalar component from its normal ordered expression,
\begin{subequations}
\begin{align}
    E_{\textrm{hf}} &= \sum_{p} z_{p}^{2} h_{pp} + \sum_{p,q} \frac{1}{2} z_{p}^{2}  z_{q}^{2}  v_{pqpq}
    \\
    &= \sum_p \frac{h_{pp}}{1 + \mathrm{e}^{\beta (\epsilon_p - \mu)}}
    \\
     &\qquad + \frac{1}{2} \sum_{p,q} \frac{v_{pqpq}}{ (1 + \mathrm{e}^{\beta (\epsilon_p - \mu)}) (1 + \mathrm{e}^{\beta (\epsilon_q - \mu)})},
    \nonumber
\end{align}
\end{subequations}
which recovers the standard thermal Hartree-Fock\cite{mermin_stability_1963,sokoloff_consequences_1967} expression for the energy.  One can find the appropriate one-electron basis or molecular orbitals by variationally minimizing the appropriate free energy i.e. the Helmholtz free energy in the canonical ensemble and the grand potential in the grand canonical ensemble.

\section{Framework for Correlated Methods
\label{sec3}}
We have noted that a practical wave function-based framework for the study of the thermal properties of electronic systems is highly desirable, and have seen that thermofield dynamics allows such a framework at the mean-field level.  Here, we wish to include correlation atop thermal mean-field.

Correlated methods frequently use a mean-field reference as the starting point, and we wish to do so here as well, but we face an additional choice which we wish to explore.  Recall that the thermal state $|\Psi(\beta)\rangle$ is obtained not from an eigenvalue problem but from an imaginary-time Schr\"{o}dinger equation.  We can choose as our reference the mean-field thermal state corresponding to a fixed temperature $\beta_0$, or we can instead use as our reference the mean-field thermal state corresponding to the temperature of interest.  By analogy with similar frameworks for coordinates in fluid dynamics as well as general relativity, we call the former approach a \textit{fixed-reference} formulation and the latter a \textit{covariant} formulation.  Mathematically, the exact thermal vacuum $\vert \Psi (\beta) \rangle$ is represented as
\begin{equation}
\left \vert \Psi (\beta) \right \rangle = \hat{\Gamma}(\beta, \beta_0) \left \vert 0(\beta_0) \right \rangle,
\end{equation}
in the fixed-reference approach, where $\hat{\Gamma}$ is a wave operator which builds correlation on the reference, and as 
\begin{equation}
\left \vert \Psi (\beta) \right \rangle = \hat{\Gamma}(\beta, \beta) \left \vert 0(\beta) \right \rangle
\end{equation}
in the covariant approach.

On the one hand, the covariant approach would seem to be more sensible, as less is demanded of the wave operator $\hat{\Gamma}$; on the other hand, the fixed-reference approach has the advantage that the quasiparticle creation and annihilation operators given by the thermal Bogoliubov transformation of Eq.~\ref{hfb-eq} are not themselves temperature-dependent, which considerably simplifies the formulation of correlated methods.  In principle any inverse temperature $\beta_0$ can be used in the fixed-reference case, but in practice the most convenient choice is $\beta_0 = 0$ for which the mean-field thermal state is exact and the wave operator $\hat{\Gamma}(0,0)$ is simply the identity operator.

Beyond deciding between the fixed-reference and covariant cases, we have a second decision to make.  The mathematical structure of the thermal averages expressed in the TFD formalism exhibits a great degree of flexibility in the way in which the \textit{bra} and the \textit{ket} thermal states can be split.  This is a direct consequence of the cyclic property of the trace involved, or equivalently the freedom in the choice of the path integral representation of TFD.\cite{das_operator_2016}

Consider the thermal average of some physical quantity $A$ which is written in TFD as
\begin{equation}
\langle A \rangle = \frac{\left\langle \mathbb{I} \right\vert \mathrm{e}^{-\beta H/2} A \mathrm{e}^{-\beta H/2} \left\vert \mathbb{I} \right\rangle}{\left\langle \mathbb{I} \right\vert \mathrm{e}^{-\beta H} \left\vert \mathbb{I} \right\rangle}.
\label{tfd-exp1}
\end{equation}
The numerator in Eq.~\ref{tfd-exp1} represents the trace of the operator $A$ along with the density operator $\hat{\rho}$. The cyclic property of the trace allows us to rewrite Eq.~\ref{tfd-exp1} as
\begin{equation}
\langle A \rangle = \frac{1}{\mathcal{Z}} \, \left\langle \mathbb{I} \right\vert \mathrm{e}^{-(1-\sigma) \beta H} \, A \, \mathrm{e}^{-\sigma \beta H} \left\vert \mathbb{I} \right\rangle,
\label{tfd-exp2}
\end{equation}
where we are free to choose $0 \leq \sigma \leq 1$.\cite{matsumoto_thermo_1983,das_operator_2016}
The cyclic property of the trace is applicable to an expectation value here because $A$ acts only on the states in $\mathcal{H}$.  Note that for $\sigma \neq 1/2$, we would require not one but \textit{two} thermal states (one each for the \textit{bra} and the \textit{ket}).  Two values of $\sigma$ are particularly useful: $\sigma = 1/2$ for which we require only a single thermal state, and $\sigma = 1$ for which one of the two requisite thermal states is simply $\langle \mathbb{I}|$.  When this $\sigma$-dependence is needed to properly interpret the equations, we will hereafter indicate it by the inclusion of a subscript which for the sake of economy we will suppress when it is not needed.

We close this section by giving reasonably explicit recipes for the thermal state in the fixed-reference and covariant cases, assuming that we evolve in $\beta$ from $\beta = 0$.  For the fixed-reference case, with $\beta_0 = 0$, we have
\begin{equation}
|\Psi_\sigma(\beta)\rangle = \hat{\Gamma}_\sigma(\beta,0) \, |\mathbb{I}\rangle,
\end{equation}
where we satisfy
\begin{subequations}
\label{fixed-correlated}
\begin{align}
\sigma \, H \, |\Psi_\sigma(\beta)\rangle 
 &= -\left(\frac{\partial \hfill}{\partial \beta} \hat{\Gamma}_\sigma(\beta,0)\right) |\mathbb{I}\rangle,
\\
\hat{\Gamma}_\sigma(0,0) &= \hat{1},
\end{align}
\end{subequations}
and where $\hat{1}$ is the identity operator.  In contrast, for the covariant case we have
\begin{equation}
|\Psi_\sigma(\beta)\rangle = \hat{\Gamma}_\sigma(\beta,\beta) \, |0_\sigma(\beta)\rangle,
\end{equation}
and we satisfy
\begin{subequations}
\label{cov-correlated}
\begin{align}
\sigma \, H_0 \, |0_\sigma(\beta)\rangle =& -\frac{\partial\hfill}{\partial\beta} |0_\sigma(\beta)\rangle,
\label{CovariantMF}
\\
\sigma \, H \, |\Psi_\sigma(\beta)\rangle =&  -\left(\frac{\partial \hfill}{\partial \beta} \hat{\Gamma}_\sigma(\beta,\beta)\right) |0_\sigma(\beta)\rangle
\label{CovariantCorr}
\\
 &- \hat{\Gamma}_\sigma(\beta,\beta) \, \frac{\partial \hfill}{\partial\beta} |0_\sigma(\beta)\rangle,
\nonumber
\\
\hat{\Gamma}_\sigma(0,0) =& \, \hat{1}.
\end{align}
\end{subequations}
Inserting Eq.~\ref{CovariantMF} into Eq.~\ref{CovariantCorr} gives us an alternative expression:
\begin{equation}
\sigma \, \bar{H}_\sigma(\beta) |0_\sigma(\beta)\rangle = -\left(\frac{\partial \hfill}{\partial \beta} \hat{\Gamma}_\sigma(\beta,\beta)\right) |0_\sigma(\beta)\rangle
\end{equation}
where the $\beta$-dependent effective Hamiltonian is
\begin{equation}
\bar{H}_\sigma(\beta) = H_0 \, \hat{\Gamma}_\sigma(\beta,\beta) -  \hat{\Gamma}_\sigma(\beta,\beta) \, H_0 + V \,  \hat{\Gamma}_\sigma(\beta,\beta)
\label{Eqn:DefHBar}
\end{equation}
and $V = H - H_0$ is the correction to the mean-field Hamiltonian $H_0$.
In Sec.~\ref{sec4}, we will show detailed formulations for these two approaches through the example of thermal configuration interaction (CI) theory.

\subsection{Chemical potential}
The framework described so far in Sec.~\ref{sec3} is common to both the canonical and the grand canonical ensemble.
However, as we have noted, working in the grand canonical ensemble is more convenient. Consequently, the thermal state, and therefore all physical quantities, are a function of two state variables:  the inverse temperature $\beta$ and chemical potential $\alpha$. 
While we have focussed our discussion on the evolution of the thermal state with respect to $\beta$, we should also discuss its evolution with respect to $\alpha$.

Just as with the $\beta$-evolution, we can evolve in $\alpha$ using either a fixed-reference or a covariant approach; further, we also have the freedom in splitting the contribution from $\alpha$ (cf. Eq.~\ref{beta-chem-th-state}) between the \textit{bra} and the \textit{ket}. In principle, one can independently choose between the fixed-reference and covariant approaches for the $\alpha$- and $\beta$-evolutions, and can have different $\sigma$'s for the $\alpha N$ and $\beta H$ terms of the $e^{\alpha N - \beta H}$. In practice, we prefer to either use the fixed-reference approach for both $\alpha$- and $\beta$-evolutions or the covariant approach for both, and we use the same $\sigma$ as well (so that we partition $\alpha N - \beta H$ rather than the two terms separately).

At $\beta=0$, we pick some chemical potential $\alpha_0$ as the initial value. Thereafter, for the fixed-reference case, we have
\begin{subequations}
    \begin{align}
        \vert \mathbb{J}\rangle &= e^{\alpha_0 N} \vert \mathbb{I} \rangle, \\
        \vert \Psi_\sigma (\alpha,\beta) \rangle &= \hat{\Gamma}_\sigma \left ( \alpha,\beta;\alpha_0,0 \right) \vert \mathbb{J} \rangle,
    \end{align}
\end{subequations}
which, in addition to Eq.~\ref{fixed-correlated} (which now uses $\vert \mathbb{J} \rangle$ instead of $\vert \mathbb{I} \rangle$ as the reference), satisfies
\begin{subequations}
    \begin{align}
        \sigma N \vert \Psi (\alpha,\beta) \rangle &= \left ( \frac{\partial}{\partial \alpha} \hat{\Gamma}_\sigma(\alpha,\beta;\alpha_0,0)  \right )\vert \mathbb{J} \rangle, \\
        \hat{\Gamma}_\sigma (\alpha,\beta;\alpha_0,0)  &= \hat{1}.
    \end{align}
\end{subequations}
On the other hand, for the covariant formulation, we have
\begin{equation}
    \vert \Psi_\sigma (\alpha,\beta) \rangle = \hat{\Gamma}_\sigma (\alpha,\beta; \alpha,\beta) \vert 0_\sigma (\alpha,\beta) \rangle
\end{equation}
which, in addition to Eq.~\ref{cov-correlated} (where the mean-field reference $\vert 0_\sigma (\alpha,\beta)\rangle$ now also depends on $\alpha$), satisfies
\begin{subequations}
    \begin{align}
        \sigma N \vert 0_\sigma (\alpha,\beta)\rangle &= \frac{\partial}{\partial \alpha} \vert 0_\sigma (\alpha,\beta) \rangle,\\
        \sigma N \vert \Psi_\sigma (\alpha,\beta) \rangle &= \left ( \frac{\partial}{\partial \alpha} \hat{\Gamma}_\sigma (\alpha,\beta;\alpha,\beta) \right ) \vert 0_\sigma(\alpha,\beta)\rangle \\
        & + \hat{\Gamma}_\sigma (\alpha,\beta;\alpha,\beta) \frac{\partial}{\partial \alpha} \vert 0_\sigma (\alpha,\beta) \rangle, \nonumber\\
        \hat{\Gamma}_\sigma (0,0;0,0) &= \hat{1}.
    \end{align}
\end{subequations}
Because we wish to ultimately make contact with the canonical ensemble at zero temperature, we select the initial chemical potential $\alpha_0$ so that the reference state $\vert \mathbb{J}\rangle$ at $\beta=0$ has the desired average particle number, and for each $\beta$ we evolve in $\alpha$ so that $\langle N \rangle $ is constant.

\section{Thermal CI Theory
\label{sec4}}
Here, we specialize both the fixed-reference and the covariant versions of our general theory to the case of thermal CI.   As will be clear, these approaches are quite similar to time-dependent CI\cite{klamroth_laser-driven_2003,huber_simulation_2005} at zero temperature. While CI is not size-extensive, a property which is highly desirable\cite{nooijen_reflections_2005} in the study of many-electron systems, we use it as an example to introduce thermofield-based quantum chemistry methods because of its simplicity. The construction of more sophisticated methods, such as thermal coupled cluster theory, is mathematically more involved and its details will be presented in a follow-up article.\cite{harsha_thermal_cc} Nevertheless, we here present preliminary numerical results of thermal CC with singles and doubles (CCSD) for the purpose of comparison with CI singles and doubles (CISD).

\subsection{Fixed-reference Thermal CI}
In this first approach, we express thermal averages with $\sigma = 1$, so that we use the asymmetric expectation value
\begin{equation}
    \langle A \rangle_\beta = \frac{1}{\mathcal{Z}} \, \langle \mathbb{J} \vert A \vert \psi(\alpha,\beta) \rangle,
\quad 
\vert \psi(\alpha,\beta) \rangle = \mathrm{e}^{\alpha N -\beta H} \vert \mathbb{J} \rangle,
\end{equation}
where the state $\vert \psi(\alpha,\beta) \rangle$ is governed by the imaginary-time Schr\"odinger equation
\begin{subequations}
    \label{fixed-imag-schro}
    \begin{align}
        H \vert \psi(\alpha,\beta)\rangle &= -\frac{\partial \hfill}{\partial \beta} \vert \psi(\alpha,\beta)\rangle
        \label{fixed-beta-schro}
        \\
        N \vert \psi(\alpha,\beta)\rangle &= \frac{\partial \hfill}{\partial \alpha} \vert \psi(\alpha,\beta)\rangle
    \end{align}
\end{subequations}
and is written as
\begin{subequations}
\label{fixed-ci-exp}
\begin{align}
|\psi(\alpha,\beta)\rangle &= e^{t_0} ( 1 + \hat{T} ) |\mathbb{J}\rangle,
\\
\hat{T} &= \sum_{pq} t_{pq} b^\dagger_p \, \tilde{b}^\dagger_q
\\
& + \frac{1}{(2!)^2} \, \sum_{pqrs} t_{pqrs} \, b^\dagger_p \, b^\dagger_q \, \tilde{b}^\dagger_s \, \tilde{b}^\dagger_r + \ldots,
\nonumber
\end{align}
\end{subequations}
where all the $\alpha$- and $\beta$-dependence is carried by the expansion coefficients. Notice that the wave operator in Eq.~\ref{fixed-ci-exp} is written in intermediate normalization with an exponential normalization constant. Since the norm of the thermal state represents the partition function which decays exponentially, an exponential parametrization provides numerical stability. The ($\alpha$- and $\beta$-independent) quasiparticle operators $b$, $\tilde{b}^\dagger$, and so on are defined by Eq.~\ref{hfb-eq} with 
\[
    w_p = 1/\sqrt{1 + n_f} \quad \textrm{and} \quad z_p = w_p \sqrt{n_f},
\]
where $n_f$ is the filling fraction at $\beta = 0$ for a given basis set. It is interesting to note that the wave operator $\hat{\Gamma}$ in Eq.~\ref{fixed-ci-exp} is composed of terms that contain equal number of \textit{tilde} and \textit{non-tilde} quasiparticle creation operators. This is because, by virtue of the thermal Bogoliubov transformation in Eq.~\ref{hfb-eq}, the difference in the total number of \textit{tilde} and \textit{non-tilde} quasiparticles is a symmetry of the Hamiltonian, i.e.
\begin{equation}
    [\mathcal{N},H ] = 0, \quad \mathcal{N} = \sum_p ( b_p^\dagger b_p - \tilde{b}_p^\dagger \tilde{b}_p)
\end{equation}
and therefore, only terms which contain an equal number of tilde and non-tilde quasiparticle operators would have a non-trivial contribution to the thermal state.

Substituting this ansatz into Eq.~\ref{fixed-imag-schro}, we obtain the following set of working equations for $\beta$-evolution:
\begin{subequations}
\begin{align}
    \frac{\partial t_0}{\partial \beta} &= -\frac{1}{\mathcal{Z}_{\mathbb{J}}} \, \langle \mathbb{J} | H \,(1 + \hat{T} ) |\mathbb{J}\rangle,
\label{fixed-ci0}
\\
\frac{\partial t_{pq}}{\partial \beta} &= -t_{pq} \frac{\partial t_0}{\partial \beta} -\frac{1}{\mathcal{Z}_{\mathbb{J}}} \, \langle \mathbb{J} | \tilde{b}_q \, b_p \, H \, (1 + \hat{T} ) |\mathbb{J}\rangle,
\label{fixed-ci1}
\\
\frac{\partial t_{pqrs}}{\partial \beta} &= -t_{pqrs} \frac{\partial t_0}{\partial \beta} -\frac{1}{\mathcal{Z}_{\mathbb{J}}} \, \langle \mathbb{J} | \tilde{b}_r \, \tilde{b}_s \, b_q \, b_p \, H \, (1 + \hat{T} ) |\mathbb{J}\rangle,
\label{fixed-ci2}
\end{align}
\end{subequations}
while for $\alpha$-evolution, we get
\begin{subequations}
\begin{align}
    \frac{\partial t_0}{\partial \alpha} &= \frac{1}{\mathcal{Z}_{\mathbb{J}}} \, \langle \mathbb{J} | N \,(1 + \hat{T} ) |\mathbb{J}\rangle,
\\
\frac{\partial t_{pq}}{\partial \alpha} &= -t_{pq} \frac{\partial t_0}{\partial \alpha} +\frac{1}{\mathcal{Z}_{\mathbb{J}}} \, \langle \mathbb{J} | \tilde{b}_q \, b_p \, N \, (1 + \hat{T} ) |\mathbb{J}\rangle,
\\
\frac{\partial t_{pqrs}}{\partial \alpha} &= -t_{pqrs} \frac{\partial t_0}{\partial \alpha} +\frac{1}{\mathcal{Z}_{\mathbb{J}}} \, \langle \mathbb{J} | \tilde{b}_r \, \tilde{b}_s \, b_q \, b_p \, N \, (1 + \hat{T} ) |\mathbb{J}\rangle,
\end{align}
\end{subequations}
and similarly for higher order terms, with the infinite-temperature partition function $\mathcal{Z}_{\mathbb{J}}$ being given by
\begin{equation}
\mathcal{Z}_{\mathbb{J}} = \langle \mathbb{J} | \mathbb{J} \rangle.
\end{equation}
We can integrate these equations from $(\alpha,\beta) = (0,0)$, with the initial conditions
\begin{equation*}
t_0 = 0; t_{pq} = 0; t_{pqrs} = 0, \ldots
\end{equation*}
to obtain the CI coefficients at any given thermodynamic coordinate $(\alpha,\beta)$.

\subsection{Covariant CI}
In the covariant formulation, we wish to work with an $\alpha$- and $\beta-$dependent reference and the expectation values are best described with the $\sigma=1/2$ formulation of TFD and take the form of a symmetric expectation value
\begin{equation}
\langle A \rangle = \frac{1}{\mathcal{Z}} \, \langle \Psi \vert A \vert \Psi \rangle,
\quad 
\vert \Psi \rangle = \mathrm{e}^{(\alpha N -\beta H)/2} \vert \mathbb{J} \rangle,
\end{equation}
where the thermal state $\vert \Psi \rangle $ is the same as the state described in Eq.~\ref{fci-vac} and is governed by the imaginary time Schr\"{o}dinger equation described in Eq.~\ref{imag-schro} and \ref{mu-schro}..

We parametrize the thermal state as
\begin{subequations}
\label{cov-ci-exp}
    \begin{align}
        |\Psi(\alpha,\beta)\rangle &= e^{s_0} ( 1 + \hat{S} ) |0(\alpha,\beta)\rangle,
        \\
        \hat{S} &= \sum_{pq} s_{pq} \, a_p^\dagger \, \tilde{a}_q^\dagger
        \\
         &+ \frac{1}{(2!)^2} \, \sum_{pqrs} s_{pqrs} \, a^\dagger_p \, a^\dagger_q \, \tilde{a}^\dagger_s \, \tilde{a}^\dagger_r + \ldots,
         \nonumber
    \end{align}
\end{subequations}
where unlike in Eq.~\ref{fixed-ci-exp} both the coefficients and the field operators are $\alpha$- and $\beta$-dependent.  Because the operators $a_p$, $\tilde{a}_p^\dagger$, and so forth are $\alpha$- and $\beta$-dependent, they have non-trivial $\alpha$ and $\beta$ derivatives.  For example, from the thermal Bogoliubov transformation of Eq.~\ref{hfb-eq}, we see that
\begin{subequations}
\begin{align}
\frac{ \partial a_p^\dagger }{\partial \beta} &= (\epsilon_p ) w_p z_p \tilde{a}_p,
\\
\frac{ \partial \tilde{a}_p^\dagger }{\partial \beta} &= -(\epsilon_p ) w_p z_p a_p,
\end{align}
\end{subequations}
and so on.

\begin{figure}[t]
\includegraphics[width=\columnwidth]{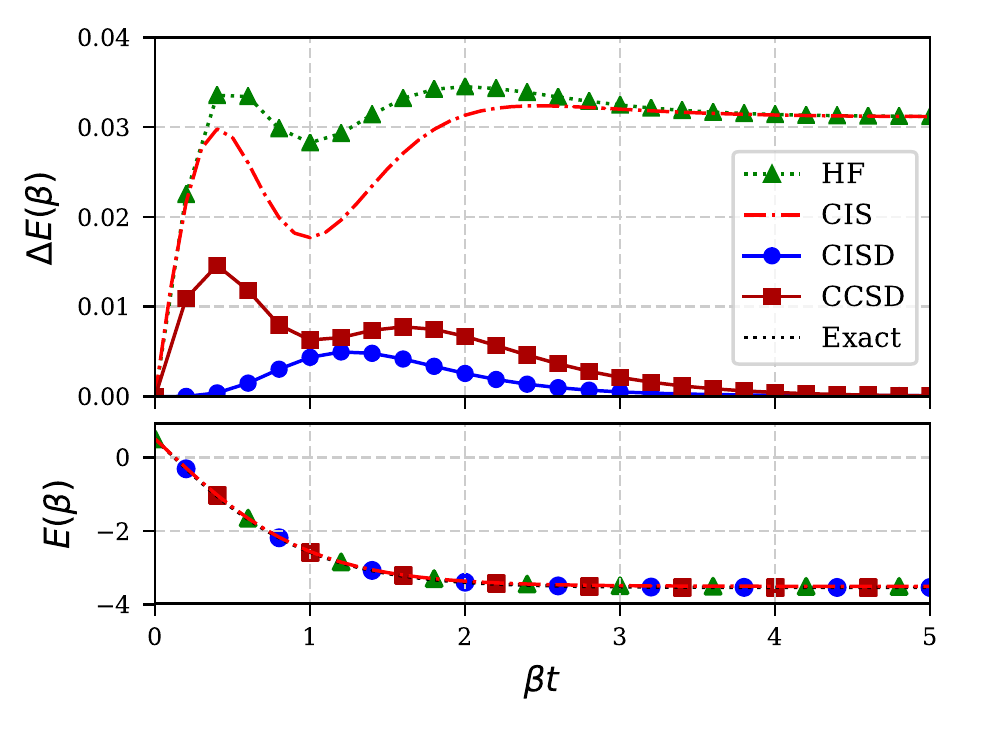}
\caption{Internal energy and error in thermal HF, covariant CIS, CISD and CCSD for the two-site Hubbard model with $U/t = 1$ at half filling on average.
\label{fig:2SitesHalfFilling}}
\end{figure}

\begin{figure}[t]
\includegraphics[width=\columnwidth]{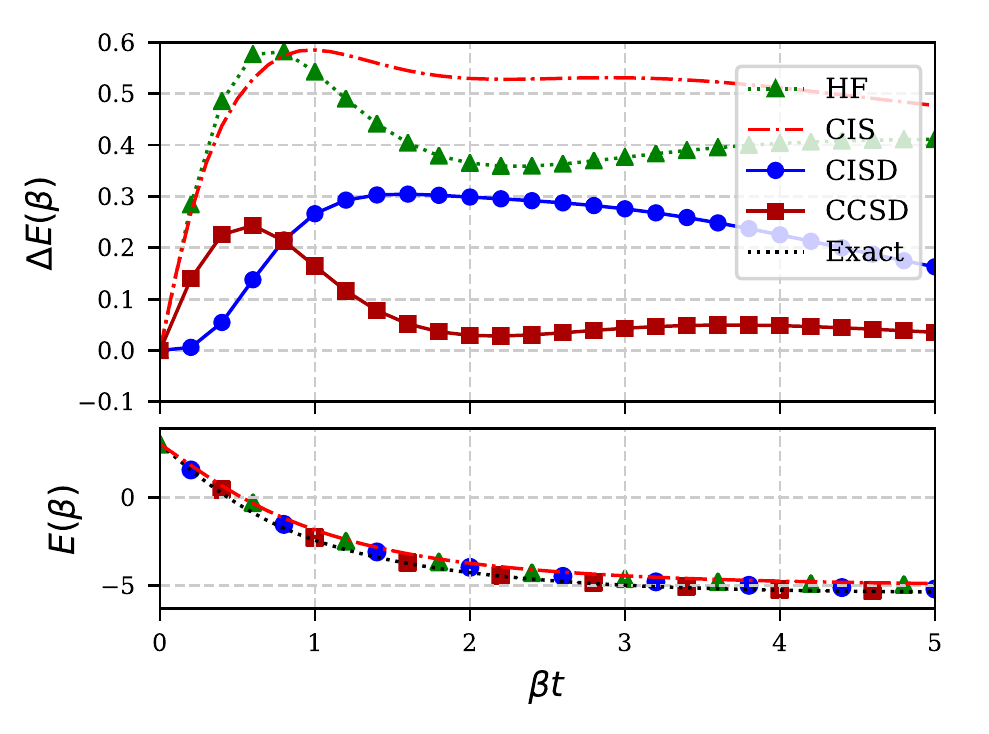}
\caption{Internal energy and error in thermal HF, covariant CIS, CISD and CCSD for the six-site Hubbard model with $U/t = 2$ at half filling on average.
\label{fig:6SitesHalfFilling}}
\end{figure}

Substituting the wave function ansatz of Eq.~\ref{cov-ci-exp} into Eq.~\ref{imag-schro}, we find
\begin{subequations}
\label{cov-ciAll-beta}
\begin{align}
    \frac{\partial s_0}{\partial \beta} &= -\frac{1}{\mathcal{Z}_0} \, \langle 0(\alpha,\beta)| \frac{1}{2} \, \bar{H} + \frac{\partial_{\mathrm{op}} \hat{S}}{\partial \beta} |0(\alpha,\beta)\rangle,
\label{cov-ci0}
\\
\frac{\partial s_{pq}}{\partial \beta} &= - s_{pq} \frac{\partial s_0}{\partial \beta} -\frac{1}{\mathcal{Z}_0} \,  \langle 0(\alpha,\beta) | \tilde{a}_q \, a_p \, \left( \frac{1}{2} \, \bar{H} \right .
\\
& \quad + \left .\frac{\partial_{\mathrm{op}} \hat{S}}{\partial \beta} \right) |0(\alpha,\beta)\rangle,
\nonumber
\label{cov-ci1} 
\end{align}
\end{subequations}
and for Eq.~\ref{mu-schro}, we get
\begin{subequations}
\label{cov-ciAll-mu}
\begin{align}
    \frac{\partial s_0}{\partial \alpha} &= \frac{1}{\mathcal{Z}_0} \, \langle 0(\alpha,\beta)| \frac{1}{2} \, \bar{N} - \frac{\partial_{\mathrm{op}} \hat{S}}{\partial \alpha} |0(\alpha,\beta)\rangle,
\\
\frac{\partial s_{pq}}{\partial \alpha} &= - s_{pq} \frac{\partial s_0}{\partial \alpha} +\frac{1}{\mathcal{Z}_0} \,  \langle 0(\alpha,\beta) | \tilde{a}_q \, a_p \, \left( \frac{1}{2} \, \bar{N} \right .
\\
& \quad - \left .\frac{\partial_{\mathrm{op}} \hat{S}}{\partial \alpha} \right) |0(\alpha,\beta)\rangle,
\nonumber
\end{align}
\end{subequations}
and so forth, where $\partial_{\mathrm{op}} \hat{S}/ \partial x$ (with $x = \alpha,\beta$) denotes the derivative of only the operator part of the wave operator $\hat{S}$, and 
\begin{subequations}
    \begin{align}
        \bar{H} &= H\, (1 + \hat{S}) - (1 + \hat{S})\, H_0
        \\
        \bar{N} &= N\, \hat{S} - \hat{S}\, N
    \end{align}
\end{subequations}
while $\mathcal{Z}_0$ is the mean-field partition function.
As with the fixed-reference case, we can integrate Eq.~\ref{cov-ciAll-beta} and \ref{cov-ciAll-mu} from $(\alpha,\beta) = (0,0)$ with the initial conditions
\begin{equation*}
    s_0 = 0; s_{pq} = 0; s_{pqrs}=0; \ldots
\end{equation*}
Our primitive implementation of both the fixed-reference and the covariant formulations of thermal CI have been made publicly accessible through a GitHub repository\cite{harsha_github} where we make use of \textit{drudge}\cite{jinmo_drudge} (computer algebra system) to generate equations and codes.

\section{Results and Conclusions
\label{sec5}}
Just as for ground-state CI, the theories described above can be truncated at various levels of excitations to yield results with respective levels of accuracy.  Here, we truncate after single (CIS) or double (CISD) excitations to study the temperature-dependence of the internal energy of the Hubbard model\cite{hubbard_electron_1963} with periodic boundary condition in the grand canonical ensemble.  The Hamiltonian for the Hubbard model is given by
\begin{equation}
H = -\frac{t}{2} \, \sum_{\langle p,q \rangle,  \sigma} \left(c_{p,\sigma}^\dagger \, c_{q,\sigma} + \textrm{h.c.}\right) + U \, \sum_{p} \hat{n}_{p,\uparrow} \, \hat{n}_{p,\downarrow},
\end{equation}
where $\langle,\rangle$ denotes that the sum is carried over sites connected in the lattice, $t$ denotes the strength of the kinetic energy term, $U$ denotes the strength of the on-site Coulomb repulsion, and $\hat{n}_{p,\sigma} = c_{p,\sigma}^\dagger \, c_{p,\sigma}$ is the number operator for lattice site $p$ and spin $\sigma$.  The correlation strength in the Hubbard model is generally characterized by the ratio $U/t$.

Recall that our formulation has been given in the grand canonical ensemble.  Our current implementation treats the inverse temperature $\beta$ and the chemical potential $\alpha$ as independent variables and evolves the differential equations of Eq.~\ref{imag-schro} and Eq.~\ref{mu-schro} in the $\alpha$-$\beta$ plane in such a way as to maintain the desired average number of particles.  We have omitted explicit orbital optimization of thermal HF at each $\alpha$, $\beta$ value, and the orbitals used in thermal HF and covariant thermal CI are all the same. We choose to work in the basis in which the Fock operator is diagonal. This particularly simplifies the thermal Bogoliubov transformation to the form described in Eq.~\ref{hfb-eq}.

\begin{figure}[t]
\includegraphics[width=\columnwidth]{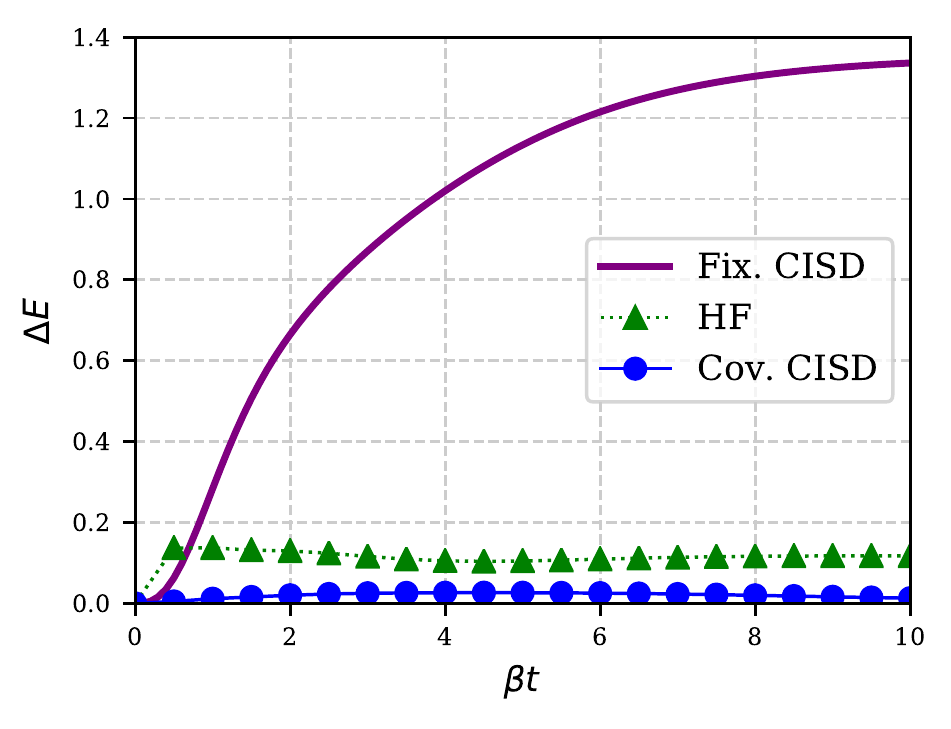}
\caption{Error in internal energy in thermal HF, covariant CISD, and fixed-reference CISD for the six-site Hubbard model with $U/t = 2$ and  two electrons on average.
\label{fig:6Sites2Electron}}
\end{figure}

We begin with the simple two-site model with (on average) two electrons and $U/t = 1$, for which both the errors in the internal energy and its value, calculated using the covariant CI method, are displayed in Fig. \ref{fig:2SitesHalfFilling}.  While thermal mean-field is exact at $\beta = 0$, we quickly see large deviations from the exact result.  Thermal CIS does not introduce any significant improvement over the Hartree-Fock internal energy, however the thermal CISD model is fairly accurate.  
Thermal CCSD further reduces the errors for moderate to large values of $\beta$, although it is not as accurate near $\beta=0$.
This is likely due to the fact that in our current implementation, thermal CC evaluates expectation values as
\[
    \langle E \rangle = \langle 0 | H | \Psi_{\textrm{CC}} \rangle
\]
while it should, in principle, include
\[
    \langle E \rangle = \langle \Psi_\textrm{L} | H | \Psi_{\textrm{CC}} \rangle,
\]
where by $\langle \Psi_\textrm{L}|$, we refer to a more elaborate, correlated \textit{bra} state.
On the other hand, thermal CI, by definition, computes thermal averages as a symmetric expectation value, with both the \textit{bra} and the \textit{ket} being correlated wavefunctions.
In the zero-temperature limit ($\beta \to \infty$) thermal HF reduces to the canonical zero-temperature ground-state restricted HF (RHF), and thermal CI reduces to the canonical zero-temperature ground-state restricted CI; for this reason, the zero-temperature limit of thermal CISD is exact (because CISD is exact for two-particle systems).  The error for intermediate values of $\beta$ with thermal CISD is because we work in the grand canonical ensemble for which the exact thermal CI requires also triple and quadruple excitations.

Figure \ref{fig:6SitesHalfFilling} shows results for the six-site model at $U/t=2$ and average half filling; again we use the covariant formulation.  The story is similar to the simpler two-site model: there are large errors at thermal HF and CIS level which are substantially reduced by thermal CI and CC.  For large $\beta$, thermal CISD reduces to the standard zero-temperature canonical ensemble CISD.

As we have seen, covariant CI works well over the entire temperature range, particularly once we include double excitations.  In contrast, the fixed-reference truncated CI does not perform well as we go away from $\beta = 0$, and soon becomes inferior to even thermal HF. This is because as we evolve away from the initial $\beta$, the mean-field contribution to the internal energy (or any other property of interest) remains fixed and all the temperature dependence and correlation effects need to be taken care of by the correlator or the wave operator. Analogous to the reference-dependent behaviour of zero temperature ground-state CI, such a method would perform poorly compared to one with a more optimized mean-field reference at each $\beta$-point.
This is emphasized in Fig.~\ref{fig:6Sites2Electron} which shows results for the six-site model at $U/t=2$ with an average of two electrons. Lastly, we note that the performance of any thermal wavefunction method is more or less equivalent to its ground-state symmetry adapted counterpart. Accordingly, the example we implement here, i.e., CI, does not perform equally well for stronger correlations, a feature shared by most ground-state methods based on the RHF reference (see, e.g., Refs. \onlinecite{henderson_quasiparticle_2014} and \onlinecite{degroote_polynomial_2016}).  Hence, we have included results only for weakly correlated Hubbard systems.

\subsection*{Conclusions}
By integrating an imaginary time Schr\"{o}dinger equation for the thermal state, thermofield theory provides an honest wave function method from which the thermal density matrix and any thermal property may be computed. This technique thus allows for the straightforward extension of ground-state methods to the computation of properties at non-zero temperatures.
Moreover, the doubling of the Hilbert space due to the introduction of the fictitious \textit{tilde} space does not affect the overall computational scaling of any method. In comparison with the ground state method, the cost increases only by a factor proportional to the number of points in the $\beta$-$\mu$ evolution.
These features make this framework ideal for application to many-body quantum systems in physics and chemistry and provide a deterministic alternative to the more common stochastic Monte-Carlo methods.  
Our early results show that even thermal CI considerably improves upon thermal mean-field theory, provided at least that one uses the covariant approach for the evolution of the thermal state. While the fixed-reference formulation is significantly easier to derive and implement, its performance is simply inadequate and we do not recommend its use.

We close by emphasizing that it was not our intention here to delve into a detailed comparison between our TFD-based approach and other finite-temperature methods. Such a comparison would be of value and, as with the details of our TFD-based coupled cluster methodology, will appear in due time. But while the story is incomplete, this early work provides both the desire and the means to generalize more sophisticated wavefunction techniques to the case of finite temperature.

\begin{acknowledgments}
    This work was supported by the U.S. Department of Energy, Office of Basic Energy Sciences, Computational and Theoretical Chemistry Program under Award No. DE-FG02-09ER16053. G.E.S. acknowledges support as a Welch Foundation Chair (No. C-0036).
\end{acknowledgments}

\bibliography{ThermoField}

\begin{thebibliography}{55}%
\makeatletter
\providecommand \@ifxundefined [1]{%
 \@ifx{#1\undefined}
}%
\providecommand \@ifnum [1]{%
 \ifnum #1\expandafter \@firstoftwo
 \else \expandafter \@secondoftwo
 \fi
}%
\providecommand \@ifx [1]{%
 \ifx #1\expandafter \@firstoftwo
 \else \expandafter \@secondoftwo
 \fi
}%
\providecommand \natexlab [1]{#1}%
\providecommand \enquote  [1]{``#1''}%
\providecommand \bibnamefont  [1]{#1}%
\providecommand \bibfnamefont [1]{#1}%
\providecommand \citenamefont [1]{#1}%
\providecommand \href@noop [0]{\@secondoftwo}%
\providecommand \href [0]{\begingroup \@sanitize@url \@href}%
\providecommand \@href[1]{\@@startlink{#1}\@@href}%
\providecommand \@@href[1]{\endgroup#1\@@endlink}%
\providecommand \@sanitize@url [0]{\catcode `\\12\catcode `\$12\catcode
  `\&12\catcode `\#12\catcode `\^12\catcode `\_12\catcode `\%12\relax}%
\providecommand \@@startlink[1]{}%
\providecommand \@@endlink[0]{}%
\providecommand \url  [0]{\begingroup\@sanitize@url \@url }%
\providecommand \@url [1]{\endgroup\@href {#1}{\urlprefix }}%
\providecommand \urlprefix  [0]{URL }%
\providecommand \Eprint [0]{\href }%
\providecommand \doibase [0]{http://dx.doi.org/}%
\providecommand \selectlanguage [0]{\@gobble}%
\providecommand \bibinfo  [0]{\@secondoftwo}%
\providecommand \bibfield  [0]{\@secondoftwo}%
\providecommand \translation [1]{[#1]}%
\providecommand \BibitemOpen [0]{}%
\providecommand \bibitemStop [0]{}%
\providecommand \bibitemNoStop [0]{.\EOS\space}%
\providecommand \EOS [0]{\spacefactor3000\relax}%
\providecommand \BibitemShut  [1]{\csname bibitem#1\endcsname}%
\let\auto@bib@innerbib\@empty
%</preamble>
\bibitem [{\citenamefont {Dirac}(1929)}]{dirac_quantum_1929}%
  \BibitemOpen
  \bibfield  {author} {\bibinfo {author} {\bibfnamefont {P.~A.~M.}\
  \bibnamefont {Dirac}},\ }\href {\doibase 10.1098/rspa.1929.0094} {\bibfield
  {journal} {\bibinfo  {journal} {P. R. Soc. London A}\ }\textbf {\bibinfo
  {volume} {123}},\ \bibinfo {pages} {714} (\bibinfo {year}
  {1929})}\BibitemShut {NoStop}%
\bibitem [{\citenamefont {Crawford}\ and\ \citenamefont
  {Schaefer}(2000)}]{crawford_introduction_2000}%
  \BibitemOpen
  \bibfield  {author} {\bibinfo {author} {\bibfnamefont {T.~D.}\ \bibnamefont
  {Crawford}}\ and\ \bibinfo {author} {\bibfnamefont {H.~F.}\ \bibnamefont
  {Schaefer}},\ }in\ \href {\doibase 10.1002/9780470125915.ch2} {\emph
  {\bibinfo {booktitle} {Reviews in {Computational} {Chemistry}}}},\ \bibinfo
  {editor} {edited by\ \bibinfo {editor} {\bibfnamefont {K.~B.}\ \bibnamefont
  {Lipkowitz}}\ and\ \bibinfo {editor} {\bibfnamefont {D.~B.}\ \bibnamefont
  {Boyd}}}\ (\bibinfo  {publisher} {John Wiley \& Sons, Inc.},\ \bibinfo {year}
  {2000})\ pp.\ \bibinfo {pages} {33--136}\BibitemShut {NoStop}%
\bibitem [{\citenamefont {Bartlett}\ and\ \citenamefont {Musia{\l
  }}(2007)}]{bartlett_coupled-cluster_2007}%
  \BibitemOpen
  \bibfield  {author} {\bibinfo {author} {\bibfnamefont {R.~J.}\ \bibnamefont
  {Bartlett}}\ and\ \bibinfo {author} {\bibfnamefont {M.}~\bibnamefont
  {Musia{\l }}},\ }\href {\doibase 10.1103/RevModPhys.79.291} {\bibfield
  {journal} {\bibinfo  {journal} {Rev. Mod. Phys.}\ }\textbf {\bibinfo {volume}
  {79}},\ \bibinfo {pages} {291} (\bibinfo {year} {2007})}\BibitemShut
  {NoStop}%
\bibitem [{\citenamefont {{\"O}stlund}\ and\ \citenamefont
  {Rommer}(1995)}]{ostlund_thermodynamic_1995}%
  \BibitemOpen
  \bibfield  {author} {\bibinfo {author} {\bibfnamefont {S.}~\bibnamefont
  {{\"O}stlund}}\ and\ \bibinfo {author} {\bibfnamefont {S.}~\bibnamefont
  {Rommer}},\ }\href {\doibase 10.1103/PhysRevLett.75.3537} {\bibfield
  {journal} {\bibinfo  {journal} {Phys. Rev. Lett.}\ }\textbf {\bibinfo
  {volume} {75}},\ \bibinfo {pages} {3537} (\bibinfo {year}
  {1995})}\BibitemShut {NoStop}%
\bibitem [{\citenamefont {Dukelsky}\ \emph {et~al.}(1998)\citenamefont
  {Dukelsky}, \citenamefont {Mart{\'i}n-Delgado}, \citenamefont {Nishino},\
  and\ \citenamefont {Sierra}}]{dukelsky_equivalence_1998}%
  \BibitemOpen
  \bibfield  {author} {\bibinfo {author} {\bibfnamefont {J.}~\bibnamefont
  {Dukelsky}}, \bibinfo {author} {\bibfnamefont {M.~A.}\ \bibnamefont
  {Mart{\'i}n-Delgado}}, \bibinfo {author} {\bibfnamefont {T.}~\bibnamefont
  {Nishino}}, \ and\ \bibinfo {author} {\bibfnamefont {G.}~\bibnamefont
  {Sierra}},\ }\href {\doibase 10.1209/epl/i1998-00381-x} {\bibfield  {journal}
  {\bibinfo  {journal} {Europhys. Lett.}\ }\textbf {\bibinfo {volume} {43}},\
  \bibinfo {pages} {457} (\bibinfo {year} {1998})}\BibitemShut {NoStop}%
\bibitem [{\citenamefont {Matsubara}(1955)}]{matsubara_new_1955}%
  \BibitemOpen
  \bibfield  {author} {\bibinfo {author} {\bibfnamefont {T.}~\bibnamefont
  {Matsubara}},\ }\href {\doibase 10.1143/PTP.14.351} {\bibfield  {journal}
  {\bibinfo  {journal} {Prog. Theor. Phys.}\ }\textbf {\bibinfo {volume}
  {14}},\ \bibinfo {pages} {351} (\bibinfo {year} {1955})}\BibitemShut
  {NoStop}%
\bibitem [{\citenamefont {Verstraete}, \citenamefont {Garc{\'i}a-Ripoll},\ and\
  \citenamefont {Cirac}(2004)}]{verstraete_matrix_2004}%
  \BibitemOpen
  \bibfield  {author} {\bibinfo {author} {\bibfnamefont {F.}~\bibnamefont
  {Verstraete}}, \bibinfo {author} {\bibfnamefont {J.~J.}\ \bibnamefont
  {Garc{\'i}a-Ripoll}}, \ and\ \bibinfo {author} {\bibfnamefont {J.~I.}\
  \bibnamefont {Cirac}},\ }\href {\doibase 10.1103/PhysRevLett.93.207204}
  {\bibfield  {journal} {\bibinfo  {journal} {Phys. Rev. Lett.}\ }\textbf
  {\bibinfo {volume} {93}},\ \bibinfo {pages} {207204} (\bibinfo {year}
  {2004})}\BibitemShut {NoStop}%
\bibitem [{\citenamefont {Feiguin}\ and\ \citenamefont
  {White}(2005)}]{feiguin_finite-temperature_2005}%
  \BibitemOpen
  \bibfield  {author} {\bibinfo {author} {\bibfnamefont {A.~E.}\ \bibnamefont
  {Feiguin}}\ and\ \bibinfo {author} {\bibfnamefont {S.~R.}\ \bibnamefont
  {White}},\ }\href {\doibase 10.1103/PhysRevB.72.220401} {\bibfield  {journal}
  {\bibinfo  {journal} {Phys. Rev. B}\ }\textbf {\bibinfo {volume} {72}},\
  \bibinfo {pages} {220401} (\bibinfo {year} {2005})}\BibitemShut {NoStop}%
\bibitem [{\citenamefont {White}(2009)}]{white_minimally_2009}%
  \BibitemOpen
  \bibfield  {author} {\bibinfo {author} {\bibfnamefont {S.~R.}\ \bibnamefont
  {White}},\ }\href {\doibase 10.1103/PhysRevLett.102.190601} {\bibfield
  {journal} {\bibinfo  {journal} {Phys. Rev. Lett.}\ }\textbf {\bibinfo
  {volume} {102}},\ \bibinfo {pages} {190601} (\bibinfo {year}
  {2009})}\BibitemShut {NoStop}%
\bibitem [{\citenamefont {Stoudenmire}\ and\ \citenamefont
  {White}(2010)}]{stoudenmire_minimally_2010}%
  \BibitemOpen
  \bibfield  {author} {\bibinfo {author} {\bibfnamefont {E.~M.}\ \bibnamefont
  {Stoudenmire}}\ and\ \bibinfo {author} {\bibfnamefont {S.~R.}\ \bibnamefont
  {White}},\ }\href {\doibase 10.1088/1367-2630/12/5/055026} {\bibfield
  {journal} {\bibinfo  {journal} {New J. Phys.}\ }\textbf {\bibinfo {volume}
  {12}},\ \bibinfo {pages} {055026} (\bibinfo {year} {2010})}\BibitemShut
  {NoStop}%
\bibitem [{\citenamefont {Pi{\v z}orn}\ \emph {et~al.}(2014)\citenamefont
  {Pi{\v z}orn}, \citenamefont {Eisler}, \citenamefont {Andergassen},\ and\
  \citenamefont {Troyer}}]{pizorn_real_2014}%
  \BibitemOpen
  \bibfield  {author} {\bibinfo {author} {\bibfnamefont {I.}~\bibnamefont
  {Pi{\v z}orn}}, \bibinfo {author} {\bibfnamefont {V.}~\bibnamefont {Eisler}},
  \bibinfo {author} {\bibfnamefont {S.}~\bibnamefont {Andergassen}}, \ and\
  \bibinfo {author} {\bibfnamefont {M.}~\bibnamefont {Troyer}},\ }\href
  {\doibase 10.1088/1367-2630/16/7/073007} {\bibfield  {journal} {\bibinfo
  {journal} {New J. Phys.}\ }\textbf {\bibinfo {volume} {16}},\ \bibinfo
  {pages} {073007} (\bibinfo {year} {2014})}\BibitemShut {NoStop}%
\bibitem [{\citenamefont {Hermes}\ and\ \citenamefont
  {Hirata}(2015)}]{hermes_finite_temperature_2015}%
  \BibitemOpen
  \bibfield  {author} {\bibinfo {author} {\bibfnamefont {M.~R.}\ \bibnamefont
  {Hermes}}\ and\ \bibinfo {author} {\bibfnamefont {S.}~\bibnamefont
  {Hirata}},\ }\href {\doibase 10.1063/1.4930024} {\bibfield  {journal}
  {\bibinfo  {journal} {J. Chem. Phys.}\ }\textbf {\bibinfo {volume} {143}},\
  \bibinfo {pages} {102818} (\bibinfo {year} {2015})}\BibitemShut {NoStop}%
\bibitem [{\citenamefont {Czarnik}, \citenamefont {Rams},\ and\ \citenamefont
  {Dziarmaga}(2016)}]{czarnik_variational_2016}%
  \BibitemOpen
  \bibfield  {author} {\bibinfo {author} {\bibfnamefont {P.}~\bibnamefont
  {Czarnik}}, \bibinfo {author} {\bibfnamefont {M.~M.}\ \bibnamefont {Rams}}, \
  and\ \bibinfo {author} {\bibfnamefont {J.}~\bibnamefont {Dziarmaga}},\ }\href
  {\doibase 10.1103/PhysRevB.94.235142} {\bibfield  {journal} {\bibinfo
  {journal} {Phys. Rev. B}\ }\textbf {\bibinfo {volume} {94}},\ \bibinfo
  {pages} {235142} (\bibinfo {year} {2016})}\BibitemShut {NoStop}%
\bibitem [{\citenamefont {Santra}\ and\ \citenamefont
  {Schirmer}(2017)}]{santra_finite-temperature_2017}%
  \BibitemOpen
  \bibfield  {author} {\bibinfo {author} {\bibfnamefont {R.}~\bibnamefont
  {Santra}}\ and\ \bibinfo {author} {\bibfnamefont {J.}~\bibnamefont
  {Schirmer}},\ }\href {\doibase 10.1016/j.chemphys.2016.08.001} {\bibfield
  {journal} {\bibinfo  {journal} {Chem. Phys.}\ }\textbf {\bibinfo {volume}
  {482}},\ \bibinfo {pages} {355} (\bibinfo {year} {2017})}\BibitemShut
  {NoStop}%
\bibitem [{\citenamefont {Zgid}\ and\ \citenamefont
  {Gull}(2017)}]{zgid_finite_2017}%
  \BibitemOpen
  \bibfield  {author} {\bibinfo {author} {\bibfnamefont {D.}~\bibnamefont
  {Zgid}}\ and\ \bibinfo {author} {\bibfnamefont {E.}~\bibnamefont {Gull}},\
  }\href {\doibase 10.1088/1367-2630/aa5d34} {\bibfield  {journal} {\bibinfo
  {journal} {New J. Phys.}\ }\textbf {\bibinfo {volume} {19}},\ \bibinfo
  {pages} {023047} (\bibinfo {year} {2017})}\BibitemShut {NoStop}%
\bibitem [{\citenamefont {Claes}\ and\ \citenamefont
  {Clark}(2017)}]{claes_finite-temperature_2017}%
  \BibitemOpen
  \bibfield  {author} {\bibinfo {author} {\bibfnamefont {J.}~\bibnamefont
  {Claes}}\ and\ \bibinfo {author} {\bibfnamefont {B.~K.}\ \bibnamefont
  {Clark}},\ }\href {\doibase 10.1103/PhysRevB.95.205109} {\bibfield  {journal}
  {\bibinfo  {journal} {Phys. Rev. B}\ }\textbf {\bibinfo {volume} {95}},\
  \bibinfo {pages} {205109} (\bibinfo {year} {2017})}\BibitemShut {NoStop}%
\bibitem [{\citenamefont {White}\ and\ \citenamefont
  {Chan}(2018)}]{white_time-dependent_2018}%
  \BibitemOpen
  \bibfield  {author} {\bibinfo {author} {\bibfnamefont {A.~F.}\ \bibnamefont
  {White}}\ and\ \bibinfo {author} {\bibfnamefont {G.~K.-L.}\ \bibnamefont
  {Chan}},\ }\href {\doibase 10.1021/acs.jctc.8b00773} {\bibfield  {journal}
  {\bibinfo  {journal} {J. Chem. Theory Comput.}\ }\textbf {\bibinfo {volume}
  {14}},\ \bibinfo {pages} {5690} (\bibinfo {year} {2018})}\BibitemShut
  {NoStop}%
\bibitem [{\citenamefont {Hummel}(2018)}]{hummel_finite_2018}%
  \BibitemOpen
  \bibfield  {author} {\bibinfo {author} {\bibfnamefont {F.}~\bibnamefont
  {Hummel}},\ }\href {\doibase 10.1021/acs.jctc.8b00793} {\bibfield  {journal}
  {\bibinfo  {journal} {J. Chem. Theory Comput.}\ }\textbf {\bibinfo {volume}
  {14}},\ \bibinfo {pages} {6505} (\bibinfo {year} {2018})}\BibitemShut
  {NoStop}%
\bibitem [{\citenamefont {Hirata}\ and\ \citenamefont
  {Jha}(2018)}]{hirata_converging_2018}%
  \BibitemOpen
  \bibfield  {author} {\bibinfo {author} {\bibfnamefont {S.}~\bibnamefont
  {Hirata}}\ and\ \bibinfo {author} {\bibfnamefont {P.~K.}\ \bibnamefont
  {Jha}},\ }\href {http://arxiv.org/abs/1812.07088} {\bibfield  {journal}
  {\bibinfo  {journal} {arXiv:1812.07088 [cond-mat, physics:physics]}\ }
  (\bibinfo {year} {2018})},\ \bibinfo {note} {arXiv: 1812.07088}\BibitemShut
  {NoStop}%
\bibitem [{\citenamefont {Sanyal}, \citenamefont {Mandal},\ and\ \citenamefont
  {Mukherjee}(1992)}]{sanyal_thermal_1992}%
  \BibitemOpen
  \bibfield  {author} {\bibinfo {author} {\bibfnamefont {G.}~\bibnamefont
  {Sanyal}}, \bibinfo {author} {\bibfnamefont {S.~H.}\ \bibnamefont {Mandal}},
  \ and\ \bibinfo {author} {\bibfnamefont {D.}~\bibnamefont {Mukherjee}},\
  }\href {\doibase 10.1016/0009-2614(92)85427-C} {\bibfield  {journal}
  {\bibinfo  {journal} {Chem. Phys. Lett.}\ }\textbf {\bibinfo {volume}
  {192}},\ \bibinfo {pages} {55} (\bibinfo {year} {1992})}\BibitemShut
  {NoStop}%
\bibitem [{\citenamefont {Sanyal}\ \emph {et~al.}(1993)\citenamefont {Sanyal},
  \citenamefont {Mandal}, \citenamefont {Guha},\ and\ \citenamefont
  {Mukherjee}}]{sanyal_systematic_1993}%
  \BibitemOpen
  \bibfield  {author} {\bibinfo {author} {\bibfnamefont {G.}~\bibnamefont
  {Sanyal}}, \bibinfo {author} {\bibfnamefont {S.~H.}\ \bibnamefont {Mandal}},
  \bibinfo {author} {\bibfnamefont {S.}~\bibnamefont {Guha}}, \ and\ \bibinfo
  {author} {\bibfnamefont {D.}~\bibnamefont {Mukherjee}},\ }\href {\doibase
  10.1103/PhysRevE.48.3373} {\bibfield  {journal} {\bibinfo  {journal} {Phys.
  Rev. E}\ }\textbf {\bibinfo {volume} {48}},\ \bibinfo {pages} {3373}
  (\bibinfo {year} {1993})}\BibitemShut {NoStop}%
\bibitem [{\citenamefont {Mandal}, \citenamefont {Sanyal},\ and\ \citenamefont
  {Mukherjee}(1998)}]{mandal_thermal_1998}%
  \BibitemOpen
  \bibfield  {author} {\bibinfo {author} {\bibfnamefont {S.~H.}\ \bibnamefont
  {Mandal}}, \bibinfo {author} {\bibfnamefont {G.}~\bibnamefont {Sanyal}}, \
  and\ \bibinfo {author} {\bibfnamefont {D.}~\bibnamefont {Mukherjee}},\ }in\
  \href@noop {} {\emph {\bibinfo {booktitle} {Microscopic {Quantum}
  {Many}-{Body} {Theories} and {Their} {Applications}}}},\ \bibinfo {series and
  number} {Lecture {Notes} in {Physics}},\ \bibinfo {editor} {edited by\
  \bibinfo {editor} {\bibfnamefont {J.}~\bibnamefont {Navarro}}\ and\ \bibinfo
  {editor} {\bibfnamefont {A.}~\bibnamefont {Polls}}}\ (\bibinfo  {publisher}
  {Springer Berlin Heidelberg},\ \bibinfo {year} {1998})\ pp.\ \bibinfo {pages}
  {93--117}\BibitemShut {NoStop}%
\bibitem [{\citenamefont {Mandal}\ \emph {et~al.}(2003)\citenamefont {Mandal},
  \citenamefont {Ghosh}, \citenamefont {Sanyal},\ and\ \citenamefont
  {Mukherjee}}]{mandal_finite_temperature_2003}%
  \BibitemOpen
  \bibfield  {author} {\bibinfo {author} {\bibfnamefont {S.~H.}\ \bibnamefont
  {Mandal}}, \bibinfo {author} {\bibfnamefont {R.}~\bibnamefont {Ghosh}},
  \bibinfo {author} {\bibfnamefont {G.}~\bibnamefont {Sanyal}}, \ and\ \bibinfo
  {author} {\bibfnamefont {D.}~\bibnamefont {Mukherjee}},\ }\href {\doibase
  10.1142/S021797920302048X} {\bibfield  {journal} {\bibinfo  {journal} {Int.
  J. Mod. Phys. B}\ }\textbf {\bibinfo {volume} {17}},\ \bibinfo {pages} {5367}
  (\bibinfo {year} {2003})}\BibitemShut {NoStop}%
\bibitem [{\citenamefont {Matsumoto}\ \emph {et~al.}(1983)\citenamefont
  {Matsumoto}, \citenamefont {Nakano}, \citenamefont {Umezawa}, \citenamefont
  {Mancini},\ and\ \citenamefont {Marinaro}}]{matsumoto_thermo_1983}%
  \BibitemOpen
  \bibfield  {author} {\bibinfo {author} {\bibfnamefont {H.}~\bibnamefont
  {Matsumoto}}, \bibinfo {author} {\bibfnamefont {Y.}~\bibnamefont {Nakano}},
  \bibinfo {author} {\bibfnamefont {H.}~\bibnamefont {Umezawa}}, \bibinfo
  {author} {\bibfnamefont {F.}~\bibnamefont {Mancini}}, \ and\ \bibinfo
  {author} {\bibfnamefont {M.}~\bibnamefont {Marinaro}},\ }\href {\doibase
  10.1143/PTP.70.599} {\bibfield  {journal} {\bibinfo  {journal} {Prog. Theor.
  Phys.}\ }\textbf {\bibinfo {volume} {70}},\ \bibinfo {pages} {599} (\bibinfo
  {year} {1983})}\BibitemShut {NoStop}%
\bibitem [{\citenamefont {Semenoff}\ and\ \citenamefont
  {Umezawa}(1983)}]{semenoff_functional_1983}%
  \BibitemOpen
  \bibfield  {author} {\bibinfo {author} {\bibfnamefont {G.~W.}\ \bibnamefont
  {Semenoff}}\ and\ \bibinfo {author} {\bibfnamefont {H.}~\bibnamefont
  {Umezawa}},\ }\href {\doibase 10.1016/0550-3213(83)90223-7} {\bibfield
  {journal} {\bibinfo  {journal} {Nucl. Phys. B}\ }\textbf {\bibinfo {volume}
  {220}},\ \bibinfo {pages} {196} (\bibinfo {year} {1983})}\BibitemShut
  {NoStop}%
\bibitem [{\citenamefont {Umezawa}(1984)}]{umezawa_methods_1984}%
  \BibitemOpen
  \bibfield  {author} {\bibinfo {author} {\bibfnamefont {H.}~\bibnamefont
  {Umezawa}},\ }\href {\doibase 10.1143/PTPS.80.26} {\bibfield  {journal}
  {\bibinfo  {journal} {Prog. Theor. Phys.}\ }\textbf {\bibinfo {volume}
  {80}},\ \bibinfo {pages} {26} (\bibinfo {year} {1984})}\BibitemShut {NoStop}%
\bibitem [{\citenamefont {Evans}\ \emph {et~al.}(1992)\citenamefont {Evans},
  \citenamefont {Hardman}, \citenamefont {Umezawa},\ and\ \citenamefont
  {Yamanaka}}]{evans_heisenberg_1992}%
  \BibitemOpen
  \bibfield  {author} {\bibinfo {author} {\bibfnamefont {T.~S.}\ \bibnamefont
  {Evans}}, \bibinfo {author} {\bibfnamefont {I.}~\bibnamefont {Hardman}},
  \bibinfo {author} {\bibfnamefont {H.}~\bibnamefont {Umezawa}}, \ and\
  \bibinfo {author} {\bibfnamefont {Y.}~\bibnamefont {Yamanaka}},\ }\href
  {\doibase 10.1063/1.529915} {\bibfield  {journal} {\bibinfo  {journal} {J.
  Math. Phys.}\ }\textbf {\bibinfo {volume} {33}},\ \bibinfo {pages} {370}
  (\bibinfo {year} {1992})}\BibitemShut {NoStop}%
\bibitem [{\citenamefont {Keldysh}(1965)}]{keldysh_diagram_1965}%
  \BibitemOpen
  \bibfield  {author} {\bibinfo {author} {\bibfnamefont {L.~P.}\ \bibnamefont
  {Keldysh}},\ }\href@noop {} {\bibfield  {journal} {\bibinfo  {journal} {Sov.
  Phys. JETP}\ }\textbf {\bibinfo {volume} {20}},\ \bibinfo {pages} {1018}
  (\bibinfo {year} {1965})}\BibitemShut {NoStop}%
\bibitem [{\citenamefont {Maldacena}(2003)}]{maldacena_eternal_2003}%
  \BibitemOpen
  \bibfield  {author} {\bibinfo {author} {\bibfnamefont {J.}~\bibnamefont
  {Maldacena}},\ }\href {\doibase 10.1088/1126-6708/2003/04/021} {\bibfield
  {journal} {\bibinfo  {journal} {J. High Energy Phys.}\ }\textbf {\bibinfo
  {volume} {2003}},\ \bibinfo {pages} {021} (\bibinfo {year}
  {2003})}\BibitemShut {NoStop}%
\bibitem [{\citenamefont {Shenker}\ and\ \citenamefont
  {Stanford}(2014)}]{shenker_multiple_2014}%
  \BibitemOpen
  \bibfield  {author} {\bibinfo {author} {\bibfnamefont {S.~H.}\ \bibnamefont
  {Shenker}}\ and\ \bibinfo {author} {\bibfnamefont {D.}~\bibnamefont
  {Stanford}},\ }\href {\doibase 10.1007/JHEP12(2014)046} {\bibfield  {journal}
  {\bibinfo  {journal} {J. High Energ. Phys.}\ }\textbf {\bibinfo {volume}
  {2014}},\ \bibinfo {pages} {46} (\bibinfo {year} {2014})}\BibitemShut
  {NoStop}%
\bibitem [{\citenamefont {Israel}(1976)}]{israel_thermo-field_1976}%
  \BibitemOpen
  \bibfield  {author} {\bibinfo {author} {\bibfnamefont {W.}~\bibnamefont
  {Israel}},\ }\href {\doibase 10.1016/0375-9601(76)90178-X} {\bibfield
  {journal} {\bibinfo  {journal} {Phys. Lett. A}\ }\textbf {\bibinfo {volume}
  {57}},\ \bibinfo {pages} {107} (\bibinfo {year} {1976})}\BibitemShut
  {NoStop}%
\bibitem [{\citenamefont {Yang}(2018)}]{yang_complexity_2018}%
  \BibitemOpen
  \bibfield  {author} {\bibinfo {author} {\bibfnamefont {R.-Q.}\ \bibnamefont
  {Yang}},\ }\href {\doibase 10.1103/PhysRevD.97.066004} {\bibfield  {journal}
  {\bibinfo  {journal} {Phys. Rev. D}\ }\textbf {\bibinfo {volume} {97}},\
  \bibinfo {pages} {066004} (\bibinfo {year} {2018})}\BibitemShut {NoStop}%
\bibitem [{\citenamefont {Suzuki}(1985)}]{suzuki_thermo_1985}%
  \BibitemOpen
  \bibfield  {author} {\bibinfo {author} {\bibfnamefont {M.}~\bibnamefont
  {Suzuki}},\ }\href {\doibase 10.1143/JPSJ.54.4483} {\bibfield  {journal}
  {\bibinfo  {journal} {J. Phys. Soc. Jpn.}\ }\textbf {\bibinfo {volume}
  {54}},\ \bibinfo {pages} {4483} (\bibinfo {year} {1985})}\BibitemShut
  {NoStop}%
\bibitem [{\citenamefont {de~Vega}\ and\ \citenamefont
  {Ba{\~n}uls}(2015)}]{de_vega_thermofield-based_2015}%
  \BibitemOpen
  \bibfield  {author} {\bibinfo {author} {\bibfnamefont {I.}~\bibnamefont
  {de~Vega}}\ and\ \bibinfo {author} {\bibfnamefont {M.-C.}\ \bibnamefont
  {Ba{\~n}uls}},\ }\href {\doibase 10.1103/PhysRevA.92.052116} {\bibfield
  {journal} {\bibinfo  {journal} {Phys. Rev. A}\ }\textbf {\bibinfo {volume}
  {92}},\ \bibinfo {pages} {052116} (\bibinfo {year} {2015})}\BibitemShut
  {NoStop}%
\bibitem [{\citenamefont {Borrelli}\ and\ \citenamefont
  {Gelin}(2016)}]{borrelli_quantum_2016}%
  \BibitemOpen
  \bibfield  {author} {\bibinfo {author} {\bibfnamefont {R.}~\bibnamefont
  {Borrelli}}\ and\ \bibinfo {author} {\bibfnamefont {M.~F.}\ \bibnamefont
  {Gelin}},\ }\href {\doibase 10.1063/1.4971211} {\bibfield  {journal}
  {\bibinfo  {journal} {J. Chem. Phys.}\ }\textbf {\bibinfo {volume} {145}},\
  \bibinfo {pages} {224101} (\bibinfo {year} {2016})}\BibitemShut {NoStop}%
\bibitem [{\citenamefont {Chen}\ and\ \citenamefont
  {Zhao}(2017)}]{chen_finite_2017}%
  \BibitemOpen
  \bibfield  {author} {\bibinfo {author} {\bibfnamefont {L.}~\bibnamefont
  {Chen}}\ and\ \bibinfo {author} {\bibfnamefont {Y.}~\bibnamefont {Zhao}},\
  }\href {\doibase 10.1063/1.5000823} {\bibfield  {journal} {\bibinfo
  {journal} {J. Chem. Phys.}\ }\textbf {\bibinfo {volume} {147}},\ \bibinfo
  {pages} {214102} (\bibinfo {year} {2017})}\BibitemShut {NoStop}%
\bibitem [{\citenamefont {Borrelli}\ and\ \citenamefont
  {Gelin}(2017)}]{borrelli_simulation_2017}%
  \BibitemOpen
  \bibfield  {author} {\bibinfo {author} {\bibfnamefont {R.}~\bibnamefont
  {Borrelli}}\ and\ \bibinfo {author} {\bibfnamefont {M.~F.}\ \bibnamefont
  {Gelin}},\ }\href {\doibase 10.1038/s41598-017-08901-2} {\bibfield  {journal}
  {\bibinfo  {journal} {Sci. Rep.}\ }\textbf {\bibinfo {volume} {7}},\ \bibinfo
  {pages} {9127} (\bibinfo {year} {2017})}\BibitemShut {NoStop}%
\bibitem [{\citenamefont {Hatsuda}(1989)}]{hatsuda_mean_1989}%
  \BibitemOpen
  \bibfield  {author} {\bibinfo {author} {\bibfnamefont {T.}~\bibnamefont
  {Hatsuda}},\ }\href
  {http://www.sciencedirect.com/science/article/pii/037594748990081X}
  {\bibfield  {journal} {\bibinfo  {journal} {Nucl. Phys. A}\ }\textbf
  {\bibinfo {volume} {492}},\ \bibinfo {pages} {187} (\bibinfo {year}
  {1989})}\BibitemShut {NoStop}%
\bibitem [{\citenamefont {Walet}\ and\ \citenamefont
  {Klein}(1990)}]{walet_thermal_1990}%
  \BibitemOpen
  \bibfield  {author} {\bibinfo {author} {\bibfnamefont {N.~R.}\ \bibnamefont
  {Walet}}\ and\ \bibinfo {author} {\bibfnamefont {A.}~\bibnamefont {Klein}},\
  }\href {\doibase 10.1016/0375-9474(90)90239-I} {\bibfield  {journal}
  {\bibinfo  {journal} {Nucl. Phys. A}\ }\textbf {\bibinfo {volume} {510}},\
  \bibinfo {pages} {261} (\bibinfo {year} {1990})}\BibitemShut {NoStop}%
\bibitem [{\citenamefont {Nocera}\ and\ \citenamefont
  {Alvarez}(2016)}]{nocera_symmetry-conserving_2016}%
  \BibitemOpen
  \bibfield  {author} {\bibinfo {author} {\bibfnamefont {A.}~\bibnamefont
  {Nocera}}\ and\ \bibinfo {author} {\bibfnamefont {G.}~\bibnamefont
  {Alvarez}},\ }\href {\doibase 10.1103/PhysRevB.93.045137} {\bibfield
  {journal} {\bibinfo  {journal} {Phys. Rev. B}\ }\textbf {\bibinfo {volume}
  {93}},\ \bibinfo {pages} {045137} (\bibinfo {year} {2016})}\BibitemShut
  {NoStop}%
\bibitem [{\citenamefont {Celeghini}\ \emph {et~al.}(1998)\citenamefont
  {Celeghini}, \citenamefont {De~Martino}, \citenamefont {De~Siena},
  \citenamefont {Iorio}, \citenamefont {Rasetti},\ and\ \citenamefont
  {Vitiello}}]{celeghini_thermo_1998}%
  \BibitemOpen
  \bibfield  {author} {\bibinfo {author} {\bibfnamefont {E.}~\bibnamefont
  {Celeghini}}, \bibinfo {author} {\bibfnamefont {S.}~\bibnamefont
  {De~Martino}}, \bibinfo {author} {\bibfnamefont {S.}~\bibnamefont
  {De~Siena}}, \bibinfo {author} {\bibfnamefont {A.}~\bibnamefont {Iorio}},
  \bibinfo {author} {\bibfnamefont {M.}~\bibnamefont {Rasetti}}, \ and\
  \bibinfo {author} {\bibfnamefont {G.}~\bibnamefont {Vitiello}},\ }\href
  {\doibase 10.1016/S0375-9601(98)00447-2} {\bibfield  {journal} {\bibinfo
  {journal} {Phys. Lett. A}\ }\textbf {\bibinfo {volume} {244}},\ \bibinfo
  {pages} {455} (\bibinfo {year} {1998})}\BibitemShut {NoStop}%
\bibitem [{\citenamefont {Floquet}, \citenamefont {Trindade},\ and\
  \citenamefont {Vianna}(2017)}]{floquet_lie_2017}%
  \BibitemOpen
  \bibfield  {author} {\bibinfo {author} {\bibfnamefont {S.}~\bibnamefont
  {Floquet}}, \bibinfo {author} {\bibfnamefont {M.~A.~S.}\ \bibnamefont
  {Trindade}}, \ and\ \bibinfo {author} {\bibfnamefont {J.~D.~M.}\ \bibnamefont
  {Vianna}},\ }\href {\doibase 10.1142/S0217751X17500154} {\bibfield  {journal}
  {\bibinfo  {journal} {Int. J. Mod. Phys. A}\ }\textbf {\bibinfo {volume}
  {32}},\ \bibinfo {pages} {1750015} (\bibinfo {year} {2017})}\BibitemShut
  {NoStop}%
\bibitem [{\citenamefont {Suzuki}(1986)}]{suzuki_thermo_1986}%
  \BibitemOpen
  \bibfield  {author} {\bibinfo {author} {\bibfnamefont {M.}~\bibnamefont
  {Suzuki}},\ }\href {\doibase 10.1007/BF01010461} {\bibfield  {journal}
  {\bibinfo  {journal} {J. Stat. Phys.}\ }\textbf {\bibinfo {volume} {42}},\
  \bibinfo {pages} {1047} (\bibinfo {year} {1986})}\BibitemShut {NoStop}%
\bibitem [{\citenamefont {Mermin}(1963)}]{mermin_stability_1963}%
  \BibitemOpen
  \bibfield  {author} {\bibinfo {author} {\bibfnamefont {N.~D.}\ \bibnamefont
  {Mermin}},\ }\href {\doibase 10.1016/0003-4916(63)90226-4} {\bibfield
  {journal} {\bibinfo  {journal} {Ann. Phys.}\ }\textbf {\bibinfo {volume}
  {21}},\ \bibinfo {pages} {99} (\bibinfo {year} {1963})}\BibitemShut {NoStop}%
\bibitem [{\citenamefont {Sokoloff}(1967)}]{sokoloff_consequences_1967}%
  \BibitemOpen
  \bibfield  {author} {\bibinfo {author} {\bibfnamefont {J.}~\bibnamefont
  {Sokoloff}},\ }\href {\doibase 10.1016/0003-4916(67)90122-4} {\bibfield
  {journal} {\bibinfo  {journal} {Ann. Phys.}\ }\textbf {\bibinfo {volume}
  {45}},\ \bibinfo {pages} {186} (\bibinfo {year} {1967})}\BibitemShut
  {NoStop}%
\bibitem [{\citenamefont {Das}\ and\ \citenamefont
  {Kalauni}(2016)}]{das_operator_2016}%
  \BibitemOpen
  \bibfield  {author} {\bibinfo {author} {\bibfnamefont {A.}~\bibnamefont
  {Das}}\ and\ \bibinfo {author} {\bibfnamefont {P.}~\bibnamefont {Kalauni}},\
  }\href {\doibase 10.1103/PhysRevD.93.125028} {\bibfield  {journal} {\bibinfo
  {journal} {Phys. Rev. D}\ }\textbf {\bibinfo {volume} {93}},\ \bibinfo
  {pages} {125028} (\bibinfo {year} {2016})}\BibitemShut {NoStop}%
\bibitem [{\citenamefont {Klamroth}(2003)}]{klamroth_laser-driven_2003}%
  \BibitemOpen
  \bibfield  {author} {\bibinfo {author} {\bibfnamefont {T.}~\bibnamefont
  {Klamroth}},\ }\href {\doibase 10.1103/PhysRevB.68.245421} {\bibfield
  {journal} {\bibinfo  {journal} {Phys. Rev. B}\ }\textbf {\bibinfo {volume}
  {68}},\ \bibinfo {pages} {245421} (\bibinfo {year} {2003})}\BibitemShut
  {NoStop}%
\bibitem [{\citenamefont {Huber}\ and\ \citenamefont
  {Klamroth}(2005)}]{huber_simulation_2005}%
  \BibitemOpen
  \bibfield  {author} {\bibinfo {author} {\bibfnamefont {C.}~\bibnamefont
  {Huber}}\ and\ \bibinfo {author} {\bibfnamefont {T.}~\bibnamefont
  {Klamroth}},\ }\href {\doibase 10.1007/s00339-004-3033-z} {\bibfield
  {journal} {\bibinfo  {journal} {Appl. Phys. A}\ }\textbf {\bibinfo {volume}
  {81}},\ \bibinfo {pages} {93} (\bibinfo {year} {2005})}\BibitemShut {NoStop}%
\bibitem [{\citenamefont {Nooijen}, \citenamefont {Shamasundar},\ and\
  \citenamefont {Mukherjee}(2005)}]{nooijen_reflections_2005}%
  \BibitemOpen
  \bibfield  {author} {\bibinfo {author} {\bibfnamefont {M.}~\bibnamefont
  {Nooijen}}, \bibinfo {author} {\bibfnamefont {K.~R.}\ \bibnamefont
  {Shamasundar}}, \ and\ \bibinfo {author} {\bibfnamefont {D.}~\bibnamefont
  {Mukherjee}},\ }\href {\doibase 10.1080/00268970500083952} {\bibfield
  {journal} {\bibinfo  {journal} {Mol. Phys.}\ }\textbf {\bibinfo {volume}
  {103}},\ \bibinfo {pages} {2277} (\bibinfo {year} {2005})}\BibitemShut
  {NoStop}%
\bibitem [{\citenamefont {Harsha}, \citenamefont {Henderson},\ and\
  \citenamefont {Scuseria}()}]{harsha_thermal_cc}%
  \BibitemOpen
  \bibfield  {author} {\bibinfo {author} {\bibfnamefont {G.}~\bibnamefont
  {Harsha}}, \bibinfo {author} {\bibfnamefont {T.~M.}\ \bibnamefont
  {Henderson}}, \ and\ \bibinfo {author} {\bibfnamefont {G.~E.}\ \bibnamefont
  {Scuseria}},\ }\href@noop {} {\bibinfo  {journal} {article in preparation}\
  }\BibitemShut {NoStop}%
\bibitem [{\citenamefont {Harsha}()}]{harsha_github}%
  \BibitemOpen
\bibfield  {journal} {  }\bibfield  {author} {\bibinfo {author} {\bibfnamefont
  {G.}~\bibnamefont {Harsha}},\ }\href
  {https://github.com/gauravharsha/thermal-ci} {\enquote {\bibinfo {title}
  {{thermalci}},}\ }\bibinfo {howpublished}
  {https://github.com/gauravharsha/thermal-ci}\BibitemShut {NoStop}%
\bibitem [{\citenamefont {Zhao}\ and\ \citenamefont
  {Scuseria}()}]{jinmo_drudge}%
  \BibitemOpen
  \bibfield  {author} {\bibinfo {author} {\bibfnamefont {J.}~\bibnamefont
  {Zhao}}\ and\ \bibinfo {author} {\bibfnamefont {G.~E.}\ \bibnamefont
  {Scuseria}},\ }\href {https://tschijnmo.github.io/drudge/} {\enquote
  {\bibinfo {title} {{drudge and gristmill}},}\ }\bibinfo {howpublished}
  {https://tschijnmo.github.io/drudge/}\BibitemShut {NoStop}%
\bibitem [{\citenamefont {{Hubbard J.}}\ and\ \citenamefont {{Flowers Brian
  Hilton}}(1963)}]{hubbard_electron_1963}%
  \BibitemOpen
  \bibfield  {author} {\bibinfo {author} {\bibnamefont {{Hubbard J.}}}\ and\
  \bibinfo {author} {\bibnamefont {{Flowers Brian Hilton}}},\ }\href {\doibase
  10.1098/rspa.1963.0204} {\bibfield  {journal} {\bibinfo  {journal} {P. R.
  Soc. Lond. A Mat.}\ }\textbf {\bibinfo {volume} {276}},\ \bibinfo {pages}
  {238} (\bibinfo {year} {1963})}\BibitemShut {NoStop}%
\bibitem [{\citenamefont {Henderson}\ \emph {et~al.}(2014)\citenamefont
  {Henderson}, \citenamefont {Scuseria}, \citenamefont {Dukelsky},
  \citenamefont {Signoracci},\ and\ \citenamefont
  {Duguet}}]{henderson_quasiparticle_2014}%
  \BibitemOpen
  \bibfield  {author} {\bibinfo {author} {\bibfnamefont {T.~M.}\ \bibnamefont
  {Henderson}}, \bibinfo {author} {\bibfnamefont {G.~E.}\ \bibnamefont
  {Scuseria}}, \bibinfo {author} {\bibfnamefont {J.}~\bibnamefont {Dukelsky}},
  \bibinfo {author} {\bibfnamefont {A.}~\bibnamefont {Signoracci}}, \ and\
  \bibinfo {author} {\bibfnamefont {T.}~\bibnamefont {Duguet}},\ }\href
  {\doibase 10.1103/PhysRevC.89.054305} {\bibfield  {journal} {\bibinfo
  {journal} {Phys. Rev. C}\ }\textbf {\bibinfo {volume} {89}},\ \bibinfo
  {pages} {054305} (\bibinfo {year} {2014})}\BibitemShut {NoStop}%
\bibitem [{\citenamefont {Degroote}\ \emph {et~al.}(2016)\citenamefont
  {Degroote}, \citenamefont {Henderson}, \citenamefont {Zhao}, \citenamefont
  {Dukelsky},\ and\ \citenamefont {Scuseria}}]{degroote_polynomial_2016}%
  \BibitemOpen
  \bibfield  {author} {\bibinfo {author} {\bibfnamefont {M.}~\bibnamefont
  {Degroote}}, \bibinfo {author} {\bibfnamefont {T.~M.}\ \bibnamefont
  {Henderson}}, \bibinfo {author} {\bibfnamefont {J.}~\bibnamefont {Zhao}},
  \bibinfo {author} {\bibfnamefont {J.}~\bibnamefont {Dukelsky}}, \ and\
  \bibinfo {author} {\bibfnamefont {G.~E.}\ \bibnamefont {Scuseria}},\ }\href
  {\doibase 10.1103/PhysRevB.93.125124} {\bibfield  {journal} {\bibinfo
  {journal} {Phys. Rev. B}\ }\textbf {\bibinfo {volume} {93}},\ \bibinfo
  {pages} {125124} (\bibinfo {year} {2016})}\BibitemShut {NoStop}%
\end{thebibliography}%
\end{document}